\documentclass[twocolumn]{bmcart}

\usepackage[utf8]{inputenc} 
\usepackage{caption}
\usepackage{multirow}
\usepackage{multicol}
\usepackage{subfigure}
\usepackage{graphicx}
\usepackage{rotating}
\usepackage{pifont}
\usepackage{amsmath}
\usepackage{bbding}
\usepackage{float} 
\usepackage{rotating}

\startlocaldefs
\endlocaldefs
\begin{document}

\begin{frontmatter}

\begin{fmbox}
\dochead{Literature Review (PhD Comprehensive Examination)}


\title{Applying Social Media Intelligence for Predicting and Identifying On-line Radicalization and Civil Unrest Oriented Threats}


\author[
 addressref={aff1},   
 corref={aff1},   
 email={swatia@iiitd.ac.in} 
]{Swati Agarwal}\\
PhD Adviser: Dr. Ashish Sureka

\address[id=aff1]{
Homepage: www.iiitd.edu.in/~swatia 
 \orgname{Indraprastha Institute of Information Technology}, 
 \street{Okhla Phase 3},   %
 \city{New Delhi},    
 \cny{India}     
 www.iiitd.ac.in 
}




\begin{abstractbox}
\begin{abstract} 
Research shows that various social media platforms on Internet such as Twitter, Tumblr (micro-blogging websites), Facebook (a popular social networking website), YouTube (largest video sharing and hosting website), Blogs and discussion forums are being misused by extremist groups for spreading their beliefs and ideologies, promoting radicalization, recruiting members and creating online virtual communities sharing a common agenda. Popular microblogging websites such as Twitter are being used as a real-time platform for information sharing and communication during planning and mobilization if civil unrest related events. Applying social media intelligence for predicting and identifying online radicalization and civil unrest oriented threats is an area that has attracted several researchers' attention over past $10$ years. There are several algorithms, techniques and tools that have been proposed in existing literature to counter and combat cyber-extremism and predicting protest related events in much advance. In this paper, we conduct a literature review of all these existing techniques and do a comprehensive analysis to understand state-of-the-art, trends and research gaps. We present a one class classification approach to collect scholarly articles targeting the topics and subtopics of our research scope. We perform characterization, classification and an in-depth meta analysis  meta-anlaysis of about $100$ conference and journal papers to gain a better understanding of existing literature.\end{abstract}




\end{abstractbox}
\end{fmbox}

\end{frontmatter}



\section{Introduction}
\begin{figure*}[t]
\centering
\begin{minipage}{.60\textwidth}
 \centering
\includegraphics[height=6cm]{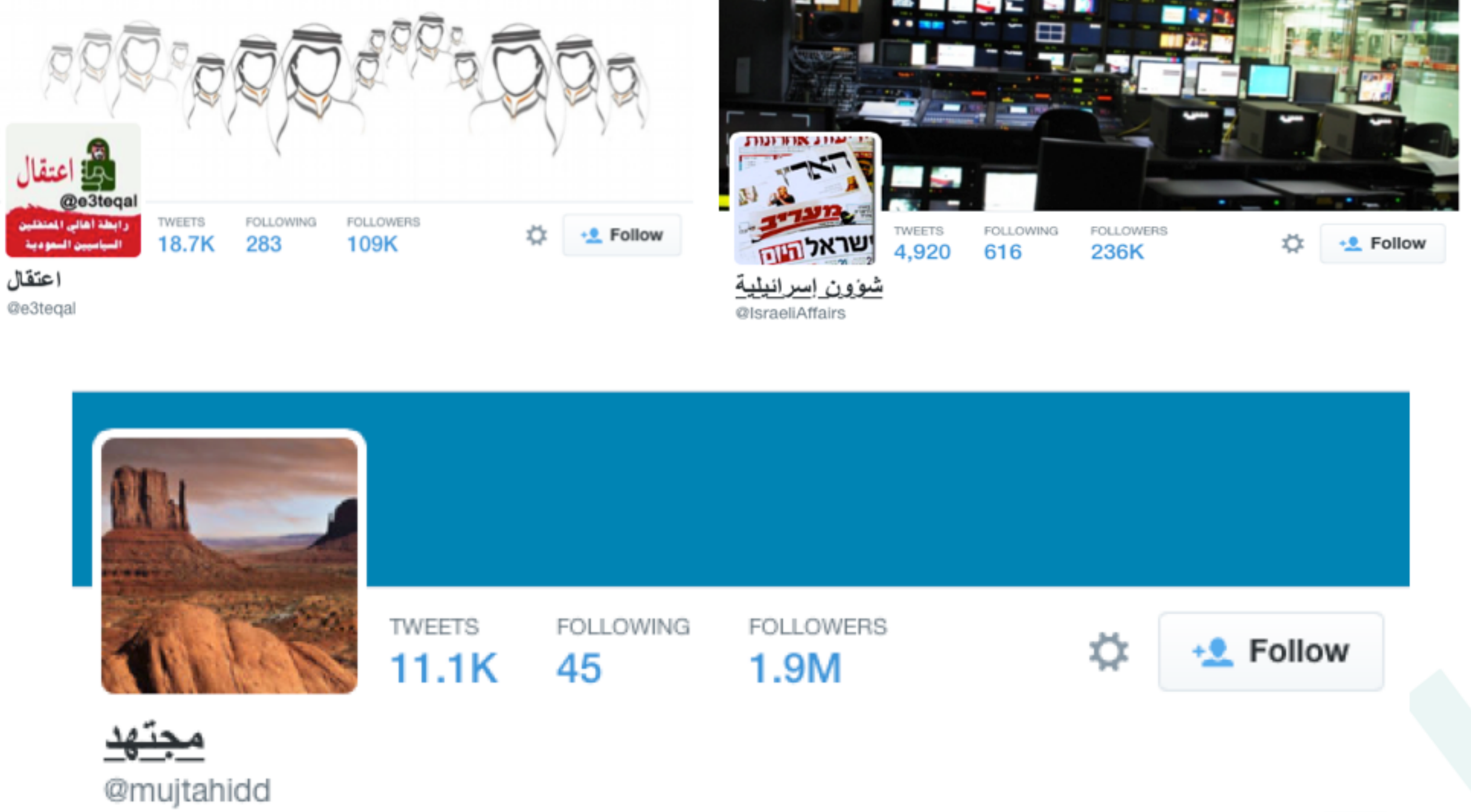}
\captionof{figure}{Top $3$ Twitter Accounts Posting Extremist Content on Website
}
\label{tech_challenge}
\end{minipage}%
\begin{minipage}{.40\textwidth}
\centering
\includegraphics[height=6cm]{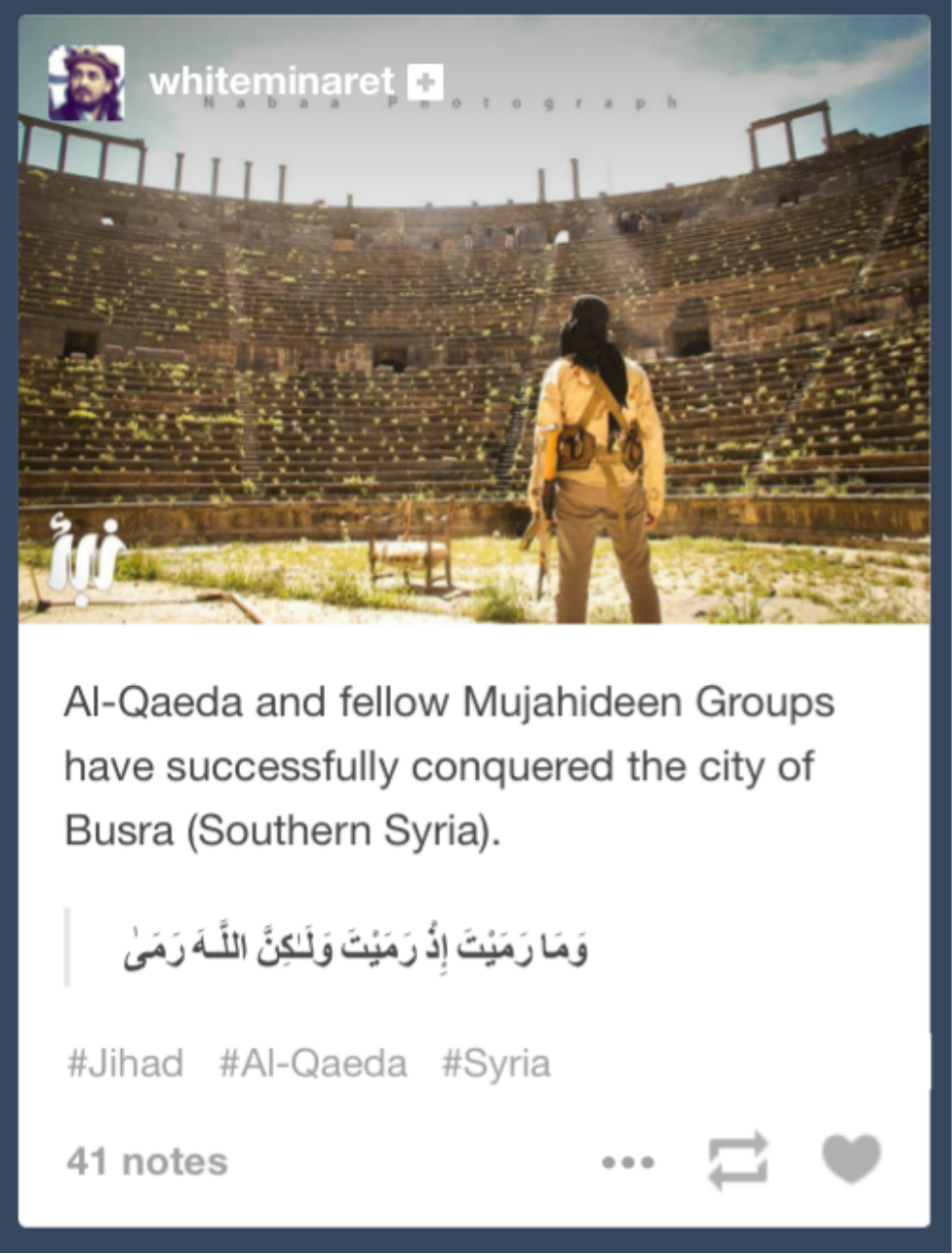}
 \captionof{figure}{An Example of a Tumblr Post Containing Multi-media Content (Image and Text) and Text Containing Multi-Lingual Script (English and Arabic)}
 \label{accounts}
\end{minipage}%
\end{figure*}

Over the past decade, social media has emerged into a dynamic form of world-wide interpersonal communication. It facilitates users for constant and continuous information sharing, making connections and conveying their thoughts across the world via different mediums. For example- social networking (Facebook), micro-blogging (Twitter, Tumblr), image sharing (Imgur, Flickr) and video hosting and sharing (YouTube, Dailymotion, Vimeo). Due to high reachability and popularity of social media websites worldwide, organizations use these websites for planning and mobilizing events for protests and public demonstrations \cite{IAAI159652}. The study of civil unrest reveals that now most of the protests are planned and mobilized in much advance \cite{Ramakrishnan:2014:BNE:2623330.2623373} \cite{Filchenkov:2014:MPS:2729104.2729135}. Crowd-buzz about these protests over social media is a rich source for civil unrest forecasting.  Traditionally, newspapers have been used a primary sources for such analysis and prediction. However, the speed and flexibility of publication on social media platforms gained the attention of various organizations for planning and making announcements of various protests, strikes, public demonstrations and riots. Similarly, simplicity of navigation, low barriers to publication (users only need to have a valid account on website) and anonymity (liberty to upload any content without revealing their real identity) have led users to misuse these websites in several ways by uploading offensive and illegal data\footnote{http://www.policechiefmagazine.org/magazine/index.cfm}. Popular social media websites, blogs, forums are frequently being misused by many hate groups to promote online radicalization (also referred as cyber-extremism, cyber-crime and cyber hate propaganda). Research shows that extremist groups put forth hateful speech, offensive and violent comments and messages focusing their mission. Many hate promoting groups use popular social media websites to promote their ideology by spreading extremist content among their viewers \cite{agarwal2014focused} \cite{mcnamee2010call}. They communicate with other such existing groups and form their virtual communities on social media sharing a common agenda. They use social networking as a medium to facilitate recruitment of new members in their group by gradually reaching world wide audiences that help to persuade others to violence and terrorism \cite{sureka2014learning}. Researchers from various disciplines like psychology, social science and computer science have constantly been developing tools and proposing techniques to counter and combat these problems of online radicalization and  generating early warning for civil unrest related events. \\
\indent Presence of extremist content and planning of civil unrest are major concerns for the government and law enforcement agencies \cite{chris2012extremism}. Online radicalization has a major impact on society that contributes to the crime against humanity and main stream morality. Presence of such content in large amount on social media is a concern for website moderators (to uphold the reputation of the website), government and law enforcement agencies (locating such users and communities to stop hate promotion and maintaining peace in country). Hence, automatic detection and analysis of radicalizing content and protest planning on social media are two of the important research problems in the domain of ISI. Monitoring the presence of such content on social media and keeping a track of this information in real time is important for security analysts who work for law enforcement agencies. Figure \ref{accounts} illustrates top $3$ Twitter accounts of hate promoting users posting extremist content very frequently and having a very large number of followers \cite{fisher2015jihadist}. According to SwarmCast journal article and Shumukh-al-islam posts, these are the three most important jihadi and support sites for Jihad and Mujahideen  Twitter.\\
\indent Over past $10$ years, since $2005$ several approaches, techniques, algorithms and tools have been proposed to mine the data and bring solutions to the problems which are encountered by Intelligence and Security Informatics (ISI)\footnote{Intelligence and security informatics is defined as an interdisciplinary research area concerned with the study of the development and use of advanced information technologies and systems for national, international, and societal security-related applications} \cite{wang2011intelligence} \cite{Chen14}. The aim of the study presented in this paper is to conduct a systematic literature survey of those previous techniques as documented in scholarly articles. Our goal is to do a comprehensive analysis of these articles to better understand the state-of-the-art, research gaps, techniques and future directions.
\subsection{Technical Challenges}
The problem of automatic identification of online radicalization (hate promoting content, users and hidden communities) and prediction of civil unrest related events (protests, riots, public demonstrations) is an important problem for ISI researchers. However, due to the dynamic nature of social media platforms, identifying such content, locating users and predicting events by keyword-based search is overwhelmingly impractical. The volume of content being posted on social media platforms makes it challenging for security analysts to discover such content manually. For example- over $6$ billion hours of video are watched each month on YouTube. $100$ millions of people perform social activities every week and millions of new subscriptions are made every day.  According to Tumblr statistics $2015$ \footnote{https://www.tumblr.com/about}, over $219$ million blogs are registered on Tumblr (second most popular micro-blogging website) and $420$ million are the active users. $80$ million posts are being published everyday, while the number of new blogs and subscriptions are $0.1$ million and $45$ thousands respectively \cite{procmicroposts2015@www2015}. Presence of large volume and multi-media content makes it difficult to find common patterns and trends in data. Security analysts also encounters problems in visualizing the data and relationship among users uploading such content . Manual inspection and keyword based flagging increase the number of false positive and reduce the efficiency of proposed approaches\\
\indent Further, textual posts on social media websites are user generated content which is unstructured and informal.  User generated data contains noisy content such as incorrect grammar, misspell words, internet slangs, abbreviations and text containing multi-lingual script. Presence of low quality content in contextual metadata increases the complexity of problem and poses technical challenges to text mining and linguistic analysis \cite{10.1371/journal.pone.0110206} \cite{naveed2011bad} \cite{korkmazchallenges}. Figure \ref{tech_challenge} shows an example of a Tumblr post that contains multi-lingual text (English and Arabic). \\
Adversarial behavior on social media also poses a technical challenge to filter the content related to hate promoting and event planning for civil disobedience. Despite providing the feasibility of posting any kind of content popular social media websites have several guidelines for uploaders posting content on the website. According to those guidelines users are not allowed to post any illegal or unethical content on the website. Due to these reasons users post information which might seem genuine but leads to sensitive information. For example, title and description of a video does not contain any suspicious term while the video has the content related to hate promotion. Similarly, a tweet might have no term related to event planning but the URL present in the tweet redirects to an external page containing information related to a public demonstration.
\subsection{Our Contributions}
This paper  falls under the aims and scope of Intelligence and Security Informatics (ISI). We have conducted an in-depth and rigorous literature survey on a sub-topic within ISI. Our literature review is comprehensive and provides insights useful to the ISI research community. To the best of our knowledge this paper is the first such survey of existing literature on social media in the domain of automated techniques for Online Radicalization detection and generating early Warning or predicting civil unrest related events. We propose a one class classification architecture across several dimensions (social media platforms being used as a data source, countering issues of online radicalization and civil unrest, machine learning techniques) to classify scholarly articles that fall under the scope of focus of our research (refer to section \ref{sec:focus}). We perform an in-depth characterization and classification based upon meta analysis of existing researches. We analyze every article and inspect commonly used techniques, identify trends and find research gaps. 
\section*{Road Map of Literature Survey}
The literature review presented in this paper is divided into multiple sections. The paper is organized as follows: Section $2$ discusses the scope and focus of our research problem where we define the topics and subtopics covered in the literature review. Section $3$ describes the general framework and the process followed to collect relevant scholarly articles for conducting literature survey. In Section $4$, we discuss the analysis of number of publications over a time period, type of paper (full, short, poster paper and journal) and their associated venues (conference, journal). In Section $5$, we perform an in-depth characterization and classification of articles based upon meta analysis (machine learning techniques, experimental data source). We also present some statistics of existing literature in various dimensions such as discriminatory features used for classification and languages being addressed in solution approach followed by the concluding remarks in Section $6$.
\begin{figure*}[ht!]
\centering{
\includegraphics[scale=0.08]{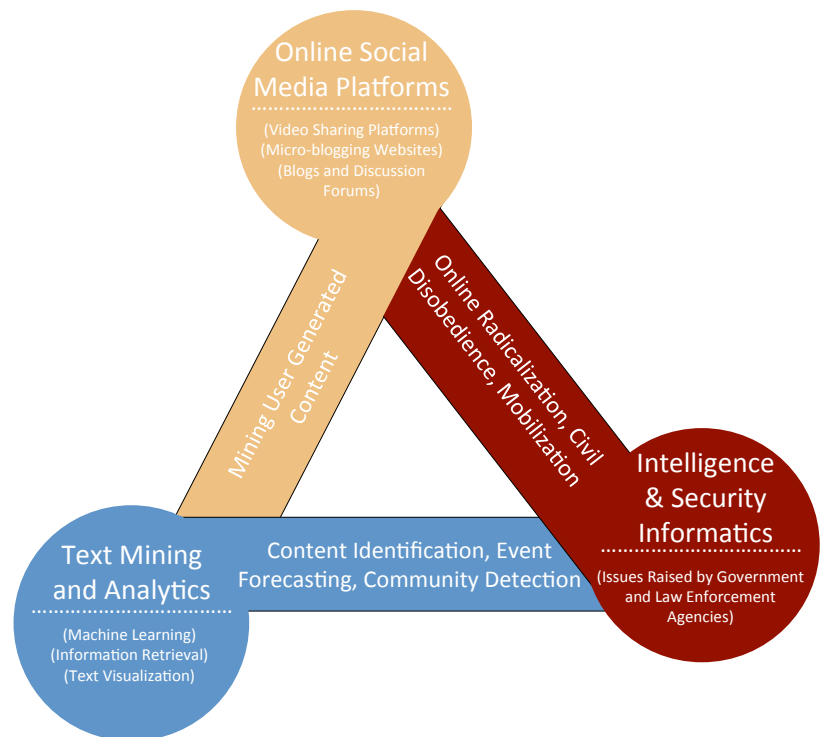}}
\caption{\label{triangle}Scope and Focus of Literature Survey: An Intersection of 3 Topics and Sub-topics (Online Social Media, Text Mining \& Analytics and Intelligence \& Security Informatics)}
\end{figure*}
\begin{figure*}
\centering{
\includegraphics[scale=0.08]{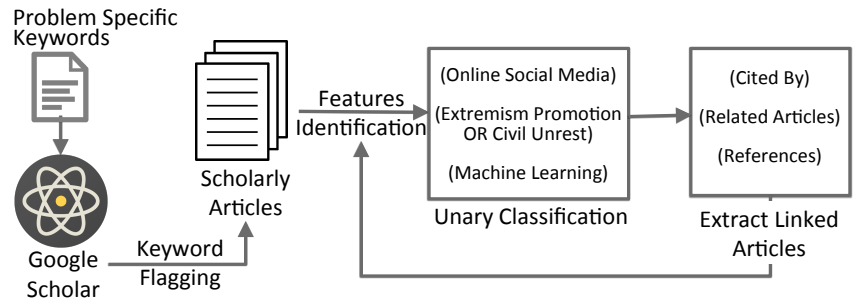}}
\caption{\label{flow_chart}General Research Framework For Literature Review Process Followed in our Survey}
\end{figure*}
\section{Research Scope and Focus}\label{sec:focus}
Figure \ref{triangle} defines the topic and scope of our literature review. As shown in Figure \ref{triangle}, our research focus is at the intersection of the following three fields:
\begin{enumerate}
\item Online Social Media Platforms
\item Intelligence and Security Informatics
\item Text Mining and Analytics
\end{enumerate}
\indent We restrict our analysis to studies on mining user generated content on social media platforms like Twitter (micro-blogging website), YouTube (video-sharing website), blogs and discussion forums and not on documents or intranet sites within an enterprise or organization. Our focus is on mining information which is publicly available (open-source intelligence). Web 2.0 and Social Media platforms contain content belonging to various modalities like image, audio, video and text. We restrict our analysis on the application of machine learning and information retrieval techniques on free-form textual data and not sound, image and video data present in social media platforms. Intelligence and security informatics is a vast field and for the literature survey presented in this paper, we focus on two important applications: online radicalization and civil unrest.
\section{General Research Framework For Literature Review}
\indent Figure \ref{flow_chart} illustrates the sequence of steps followed by us to collect papers for conducting our literature review. As shown in Figure \ref{flow_chart}, we begin by creating a list of key-terms representing our topic or problem area. For example, some of the search key-terms to retrieve relevant papers are: 'civil unrest/protest', 'event forecasting', 'early warning', 'extremism detection', 'online radicalizing community detection'. We search relevant articles using Google Scholar\footnote{https://scholar.google.co.in/} which is a well-known and widely used web-based search engine for finding scholarly literature. We use Google Scholar as it has a good coverage and index of articles and also provides a powerful mechanism to explore related work, citations and authors. Google Scholar also provides data on how often and how recently a paper has been cited which is also a useful metric (judging the impact) for conducting our literature survey. We go through the Title, Keywords and Abstract of the articles retrieved as a result of key-term based search as well as through related article and cited by links provided for each article in the search result. We perform a meta analysis on article and determine the relevance of the article to our topic or focus area. We perform a one class classification (Rule Based Classifier) on each article and check if it meets the scope and focus defined in Section \ref{sec:focus}. During classification, we typically mine the Title, Keywords and Abstract and accordingly save the article in our database of collection of closely related work. 
\begin{figure*}[ht!]
\centering
\includegraphics[width=0.97\textwidth]{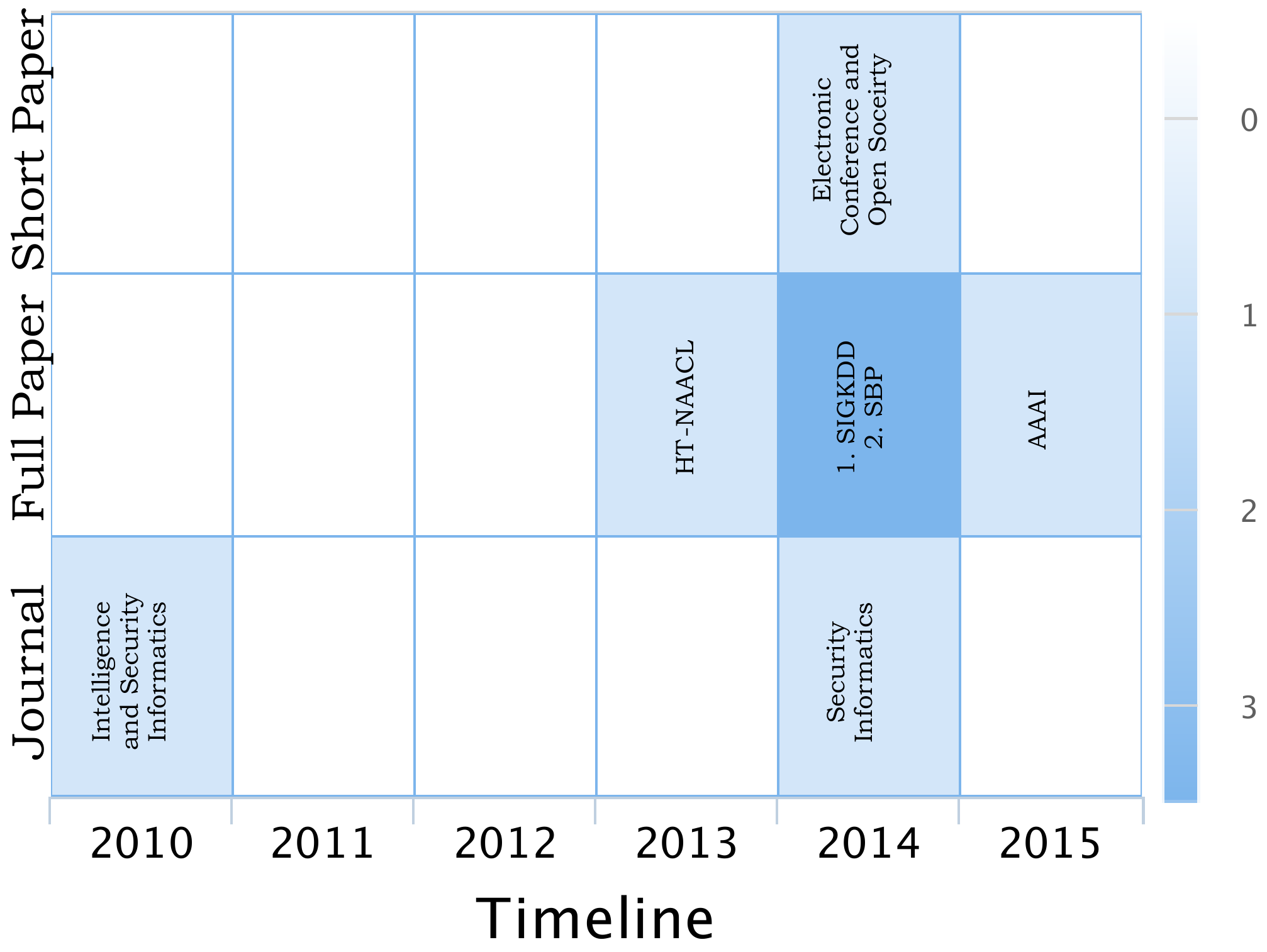}
\caption{A Heatmap Representation of the Variance in Number of Publications and Conference Venues [Civil Unrest Related Event Forecasting] Over a Time Period}
\label{heatmap_unrest}
\end{figure*}
\begin{figure*}
\centering
\includegraphics[width=1\textwidth]{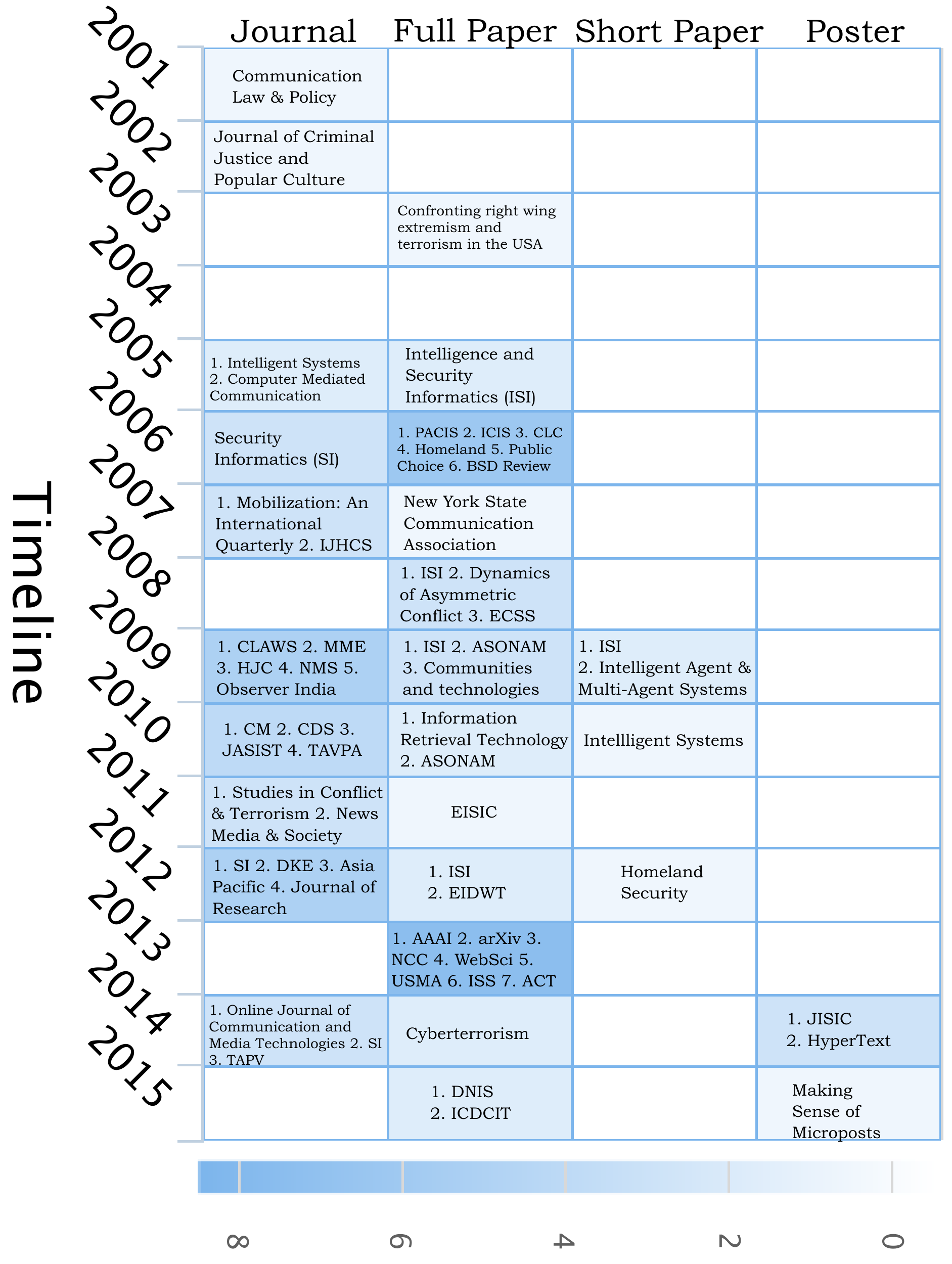}
\caption{A Heatmap Representation of the Variance in Number of Publications and Conference Venues [Online Radicalization] Over a Time Period}
\label{heatmap_radicalization}
\end{figure*}
\section{Survey of Number of Publications and Venue}\label{analysis_venue}
\indent We limit our literature review to papers published in peer-reviewed conference and workshop proceedings and journals (and indexed in bibliographic databases such as IEEE XPlore, ACM Digital Library and Springer). We do not include granted patents or patent applications, books, lecture notes or slides, software products and information websites. We record the year, venue (conference or journal name) and type of every paper (full, short or poster paper in-case of a conference). As shown in Figures \ref{heatmap_unrest} and \ref{heatmap_radicalization}, we classify each paper (civil unrest and online radicalization) in terms of its type (y-axis), year (x-axis) and the venue (value in the cell). The graphs in Figures \ref{heatmap_unrest} and \ref{heatmap_radicalization}, also shows a gradient (gradual color change) in which each cell is background shaded with a color. The color in the cell represents the number of papers published with the period and paper type represented by the cell. Usage of color ramp or progression as one dimension helps us in understanding the trends (growing, stable, and declining) in paper publications in our focus area.
\subsection{Event Forecasting or Early Warning for Civil Unrest Related Events}
\indent Figure \ref{heatmap_unrest} reveals that there are a total of $8$ publications in both conferences and journals in the area of civil unrest forecasting by mining social media content. Figure \ref{heatmap_unrest} also reveals that maximum number of papers are published in $2014$ including one journal \cite{compton2014using}, $3$ full papers \cite{Ramakrishnan:2014:BNE:2623330.2623373} \cite{xu2014civil} \cite{Chen:2014:NSS:2623330.2623619} and one short paper \cite{Filchenkov:2014:MPS:2729104.2729135}. During the time period of $2010$ to $2013$, we only find $2$ publications one every year while there is no publication in $2011$ and $2012$ \cite{6689273} \cite{colbaugh2012early}. We observe that over past $6$ years, no conference or journal have more than one publication on this research problem except SIGKDD (has $2$ full papers in $2014$) \cite{Ramakrishnan:2014:BNE:2623330.2623373} \cite{Chen:2014:NSS:2623330.2623619}. By Figure \ref{heatmap_unrest}, we conclude that civil disobedience event forecasting problem has recently gained the attention of researchers.
\begin{figure*}
\centering{
\includegraphics[scale=0.5]{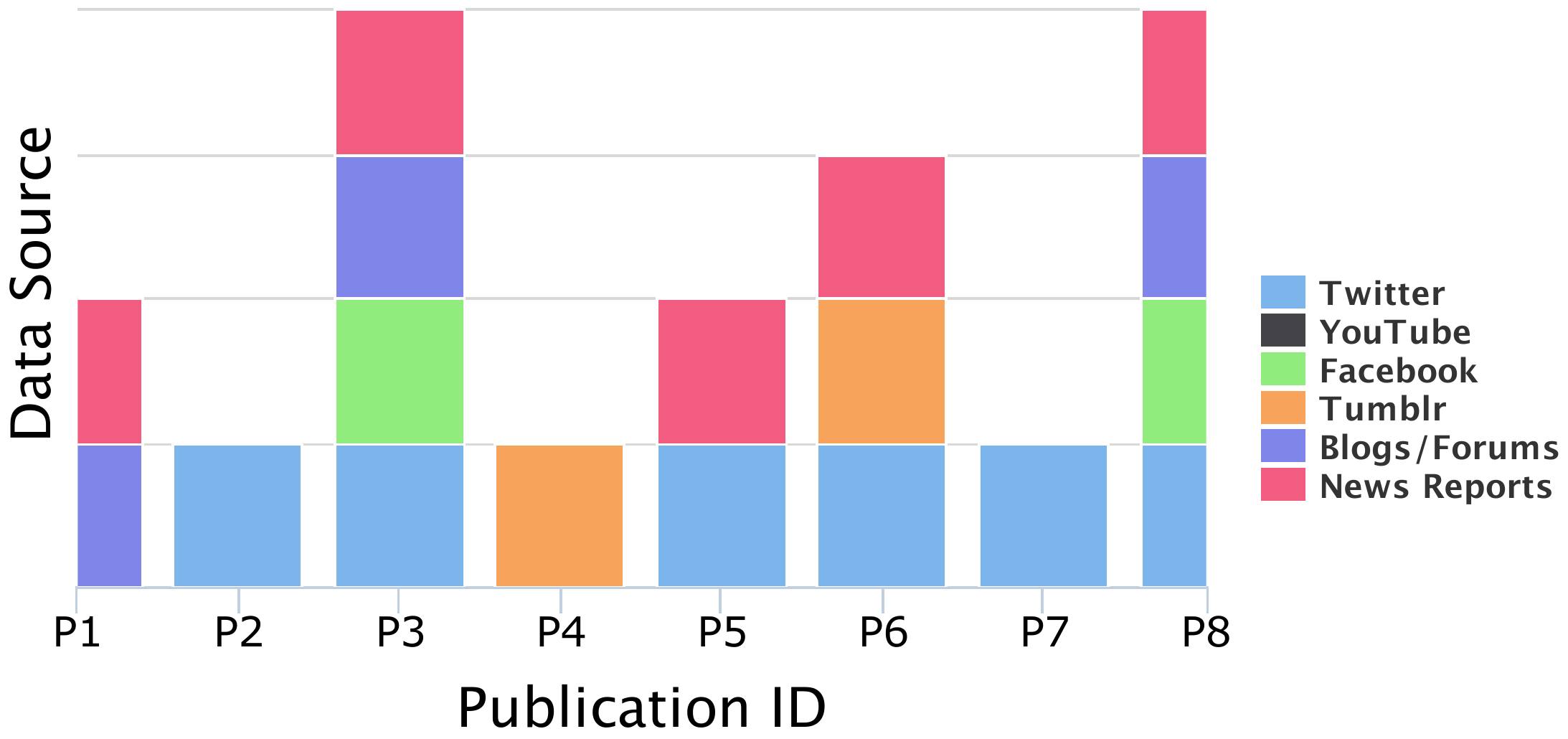}
\caption{\label{data_unrest}Different Type of Social Media Platforms Used as Data Source for Predicting Civil Unrest Related Events}}
\end{figure*}
\subsection{Identification of Online Extremist Content, Users and Communities }
\indent Similar to Figure \ref{heatmap_unrest}, Figure \ref{heatmap_radicalization} shows the number of publications and venue for existing literature conducted in the domain of online radicalization detection. Figure \ref{heatmap_radicalization} reveals that there has been a lot of work in the domain of hate and extremism detection on social media. There are many journal papers and conference proceedings available over past $15$ years. We observe that this problem is not very new. However, this problem is important and not fully solved. Researchers have constantly been working on this problem and publishing papers in reputed conferences and journals. Here, the first few publications during the time period of $2001$ to $2005$, are not using social media as a platform for conducting experiments and have been published in crime and law specific venues (both international and national) \cite{leets2001responses} \cite{schafer2002spinning}. For example, Confronting right wing extremism and terrorism in the USA (conference 2003) is a national conference exclusively for researches in the domain of hate and terrorism detection \cite{michael2003confronting}. Figure \ref{heatmap_radicalization} also reveals that papers in online radicalization detection domain are published in wide range of conferences ($32$) and journals ($23$). Maximum number of papers are published in ISI conference\footnote{http://www.eisic.eu/} (more than $6$) and Security Informatics journal\footnote{http://www.security-informatics.com/} (more than $4$) which are two of the most core venues for the research domain and problem \cite{salem2006content} \cite{yang2006analyzing} \cite{zhou2006exploring} \cite{Zhang:2009:DWF:1706428.1706441} \cite{glass2012estimating} \cite{scanlon2014automatic}. Figure \ref{heatmap_radicalization} shows that the maximum number of journal and total (including full and short papers) publications are in $2009$ \cite{Zhang:2009:DWF:1706428.1706441} \cite{bermingham2009combining} \cite{5137295} \cite{anand2009terror}. While, maximum number of conference regular papers are published in 2013 ($7$ full papers) \cite{kwok2013locate} \cite{wadhwa2013tracking} \cite{o2013uncovering} \cite{o2013analysis} \cite{berger2013matters} \cite{botha2013assessing} \cite{blanquart2013twitter} and 2006 ($6$ full papers) \cite{zhou2006exploring} \cite{chau2006framework} \cite{xu2006mining} \cite{ressler2006social} \cite{davis2006ending}. There are only $3$ poster papers published in $2014$ (JISIC \cite{sureka2014learning} and HT \cite{agarwal2014focused}) and $2015$ (Making Sense of Microposts \cite{procmicroposts2015@www2015}). Heatmap shown in Figure \ref{heatmap_radicalization} also reveals that there is no publication in $2004$.
\section{Characterization and Classification of Articles Based Upon Meta Analysis}
\indent In this part of literature review, we present a characterization based study on previous researches done in the area of civil unrest related event forecasting (refer to Table \ref{label_unrest}) and online radicalization detection (refer to Table \ref{label_radicalization}). We analyze each paper and create a list of all dimensions to demonstrate the statistics. We also present statistics of existing literature that use social media platforms as data source for conducting experiments. Each dimension is further classified in sub-categories and have their properties associated with them. These dimensions are as follows:
\begin{enumerate}
\item Data Source
\item Techniques
\item Features
\item Evaluation
\item Type of Analysis 
\item Language
\item Genre
\item Region (only for online radicalization)
\end{enumerate}
\subsection{Social Media Websites as Data Source}
\indent As discussed in Section \ref{analysis_venue}, there are only $8$ existing publications in the domain of civil disobedience event forecasting and more than $40$ publications in the area of online extremism detection. Therefore, for the problem of event forecasting for civil unrest related events, we take publication IDs on x-axis and data source on y-axis; colors represent the different social media platform and other data sources. Figure \ref{data_unrest} reveals that Twitter and News media reports are the most widely used data sources for event forecasting \cite{IAAI159652} \cite{Ramakrishnan:2014:BNE:2623330.2623373} \cite{6689273}. Apart from the news media articles, micro-blogging websites are a rich source of information. In Figure \ref{data_unrest}, we observe that the papers using a single source of data are using micro-blogging websites (Twitter and Tumblr) \cite{Filchenkov:2014:MPS:2729104.2729135} \cite{xu2014civil} \cite{6689273}. Most of the researchers have used multiple data sources (blogs, forums, news media and other social networking websites) for information extraction \cite{IAAI159652} \cite{Ramakrishnan:2014:BNE:2623330.2623373}. We also observe that despite being most popular video hosting and sharing website and widely used YouTube has not been used in any of the existing research for protest planning or prediction.\\ 
\begin{table*}[t]
\centering
\footnotesize
\renewcommand{\arraystretch}{1.25}
\caption{Statistics of Number of Publications Using Various Social Media Platforms as Data Source for Conducting Experiments over Past Decade}
\label{stats_data_radicalization}
\begin{tabular}{p{0.15\textwidth}cccccccccc}
\hline
\multirow{2}{*}{\textbf{Data Source}}&\multicolumn{10}{c}{\textbf{Years}}\\
&\textbf{2006}&\textbf{2007}&\textbf{2008}&\textbf{2009}&\textbf{2010}&\textbf{2011}&\textbf{2012}&\textbf{2013}&\textbf{2014}&\textbf{2015}\\
\hline
\textbf{Twitter}&0&0&0&1&0&0&6&7&1&1\\
\textbf{YouTube}&0&0&2&3&2&2&2&2&1&1\\
\textbf{Facebook}&0&0&0&1&0&0&3&2&0&0\\
\textbf{Tumblr}&0&0&0&0&0&0&0&0&0&1\\
\textbf{Blogs}&2&2&1&1&0&0&0&0&0&0\\
\textbf{Forums}&2&0&0&2&1&1&0&1&1&0\\
\textbf{Webpages}&2&0&0&1&0&0&0&0&0&0\\
\textbf{News Articles}&1&0&0&0&0&0&0&0&0&0\\
\textbf{Image Hosting}&0&0&0&0&0&0&0&1&0&0\\
\textbf{Gaming/Virtual World}&0&0&1&0&0&0&0&0&0&0\\
\hline
\end{tabular}
\end{table*}
\begin{figure*}[t]
\centering{
\includegraphics[scale=0.5]{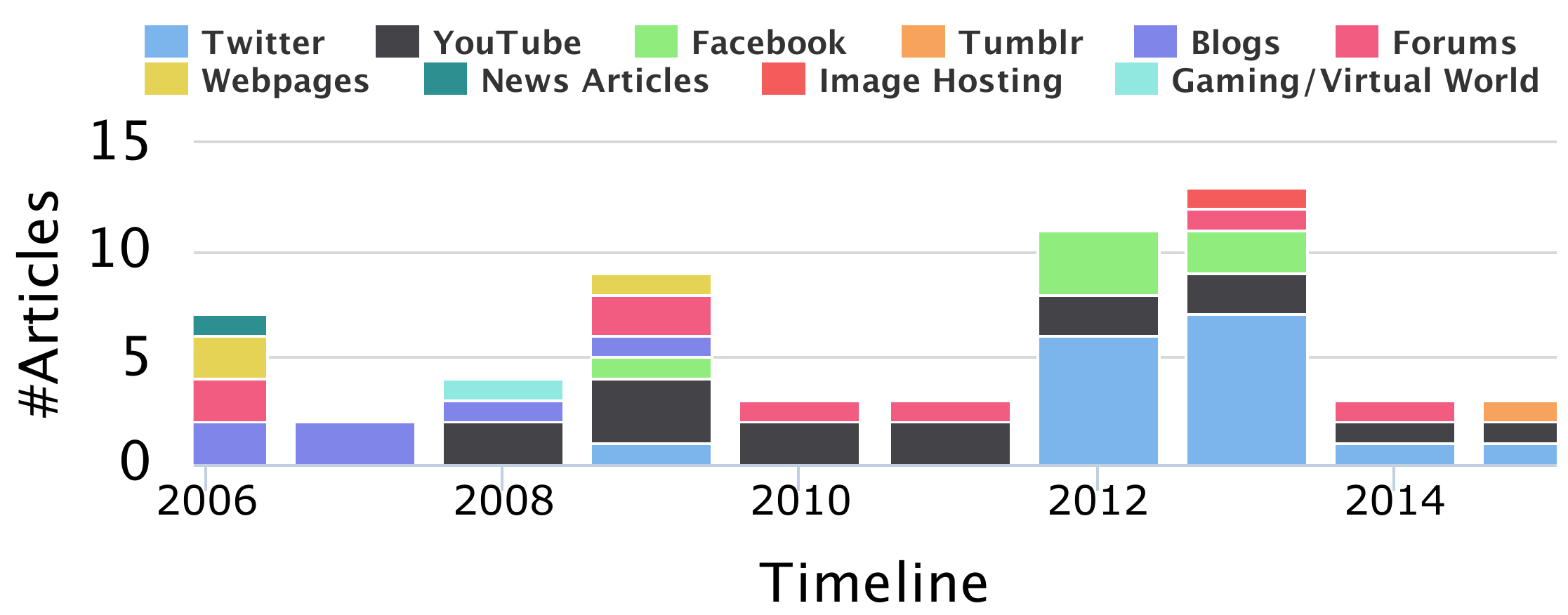}
\caption{\label{data_radicalization}Number of Publications Over a Period of Time using One or More Social Media Platforms For Online Radicalization Detection}}
\end{figure*}
\indent For online radicalization detection, we identify the number of publications (y-axis) using various data sources (legends) and experimental dataset over past decade (x-axis). Figure \ref{data_radicalization} reveals that there has been some work on all popular social networking platforms, micro-blogging websites \cite{Filchenkov:2014:MPS:2729104.2729135} \cite{xu2014civil} \cite{6689273}, video sharing \cite{5137295} \cite{Ortiz08} \cite{surekaAIRS} \cite{van2010performing} and image hosting websites \cite{weimann2014new}. Many researchers have used blogs \cite{chau2006framework} \cite{chau2007mining}, forums \cite{Zhang:2009:DWF:1706428.1706441} \cite{fu2010focused}, web documents \cite{fu2010focused}, news articles \cite{6061224} and online gaming websites \cite{chen2008cyber} for hate and extremism detection (refer to Table \ref{stats_data_radicalization} for statistics). Figure \ref{data_radicalization} reveals that initially in $2006$, most of the researches were conducted using blogs, discussion forums, news articles and extremist websites as a data source. YouTube, Facebook and Twitter were founded in $2004$, $2005$ and $2006$ respectively which gradually became popular among extremist users for posting hate promoting content. Previous research shows that YouTube is most widely used platform for hate promotion. Figure \ref{data_radicalization} also reveals YouTube is being used as a rich source of extremist data for online radicalization detection over past $7$ years \cite{5137295} \cite{Ortiz08} \cite{surekaAIRS} \cite{van2010performing} and image hosting websites \cite{weimann2014new}. Micro-blogging websites are very popular and widely used platforms among its users (including terrorist organization and hate promoting groups) for conveying messages and forming communities. Figure \ref{data_radicalization} shows that recently, Twitter has gained the attention of researchers and we find many publications in past $3$ years that are using Twitter for identifying extremism content, users and communities \cite{sureka2014learning} \cite{o2013analysis} \cite{agarwal2015using} \cite{6284099}. We also observe that despite being second most popular micro-blogging website, we find only publication that is using Tumblr (founded in $2007$) as a data source for extremism content identification \cite{procmicroposts2015@www2015}. Table \ref{stats_data_radicalization} shows that there are $15$ papers that have used YouTube as a data source and similarly $16$ papers using Twitter, $8$ papers using discussion forums and $6$ papers using blogs for collecting experimental dataset to counter and combat online radicalization.
\begin{figure*}
\centering{
\includegraphics[scale=0.12]{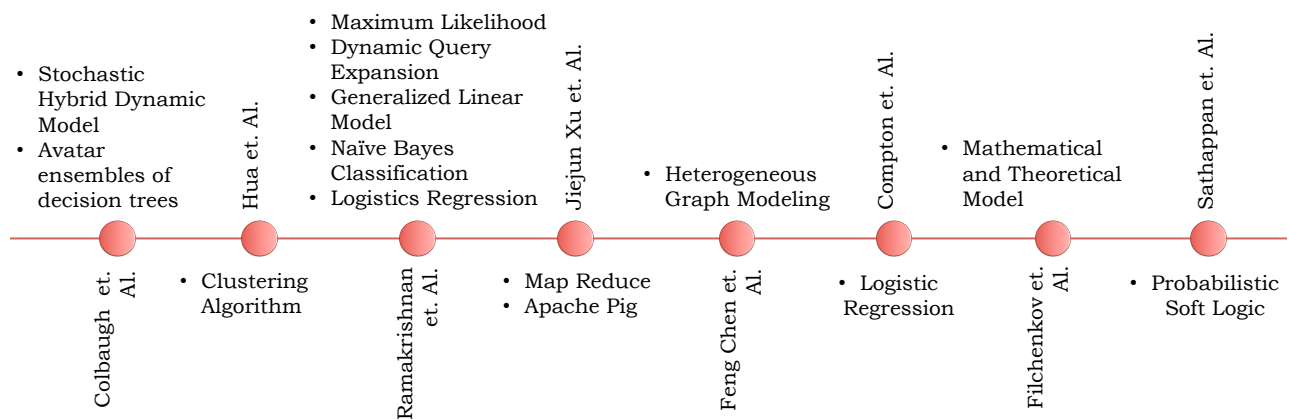}
\caption{\label{techniques_unrest}Machine Learning Techniques Used By Researchers in Existing Literature of Event Forecasting in the Domain of Civil Disobedience}
}
\end{figure*}
\subsection{Machine Learning and Data Mining Techniques} \label{Section-datamining-techniques}
\begin{figure*}[t]
\centering
\includegraphics[scale=0.12]{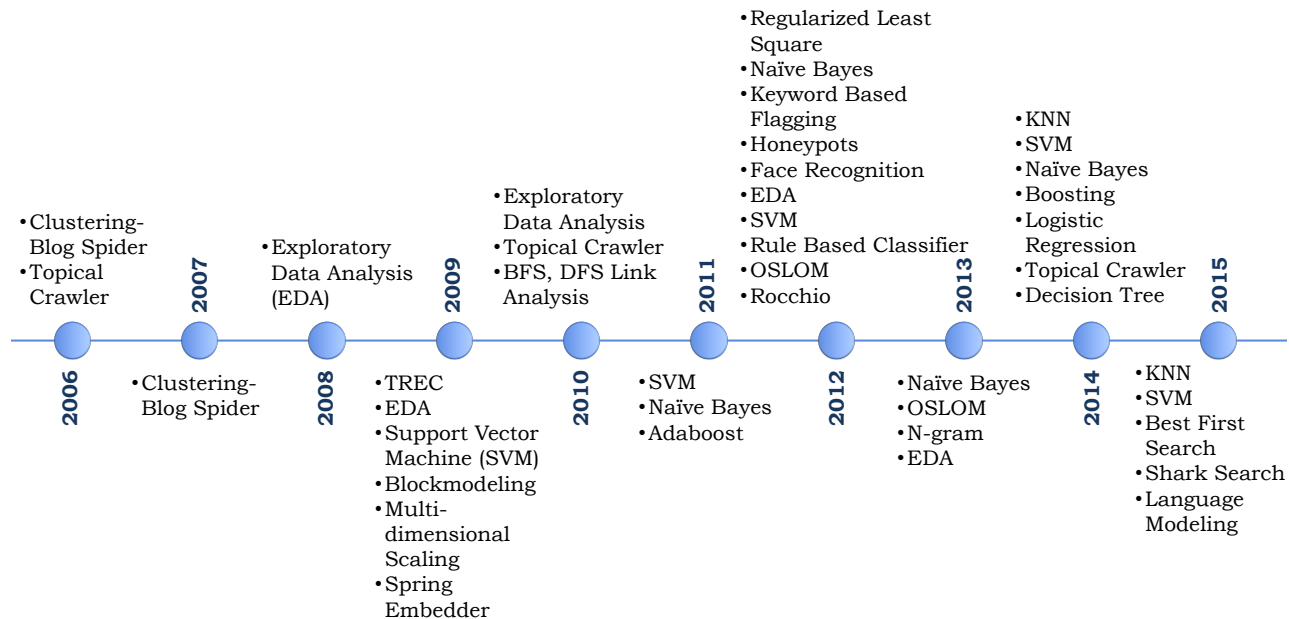}
\caption{Machine Learning Techniques Used By Researchers in Existing Literature of Online Radicalization and Hate Promotion Detection}
\label{techniques_radicalization}
\end{figure*}

Figures \ref{techniques_unrest} and \ref{techniques_radicalization} illustrate the machine learning and information retrieval techniques used in previous papers of civil unrest event prediction and online radicalization respectively.\\
\indent Similar to Figure \ref{data_unrest}, Figure \ref{techniques_unrest} illustrates the proposed methods and techniques used in each paper individually. Ramakrishnan \cite{Ramakrishnan:2014:BNE:2623330.2623373} proposed five different techniques for event forecasting for five different kinds of data and models (volume based, opinion based, tracking of activities, distribution of events and cause of protest). Unlike Ramakrishnan \cite{Ramakrishnan:2014:BNE:2623330.2623373}, in Figure \ref{techniques_unrest}, we find that most of the researchers have applied ensemble learning on multiple data mining and machine learning techniques to achieve better accuracy in prediction \cite{IAAI159652} \cite{xu2014civil} \cite{colbaugh2012early}.\\ 
\indent Clustering, Logistic Regression and Dynamic Query Expansion are the commonly used techniques to predict upcoming events related to civil unrest or protest. We go through the methodology Section of each paper and observe that named entity recognition is a common phase for all approaches illustrated in Figure \ref{techniques_unrest}. In named entity recognition, they extract several entities present in the contextual metadata (tweets, Facebook comment, News article) \cite{ritter2011named}. These entities could be spatiotemporal expressions, topic being discussed in posts (refer to Table \ref{label_unrest} for full list of entities and features). List of all entities and keywords is dynamically expanded by using Dynamic Query Expansion method where they find similar and relevant keywords/entities using external lexical source (for example- WordNet\footnote{https://wordnet.princeton.edu}, VerbOcean\footnote{http://demo.patrickpantel.com/demos/verbocean/}). Dynamic Query Expansion is an iterative process and converges once the keywords are stable. They further perform several clustering and classification techniques on these entities and text to predict upcoming events.\\
\indent Another popular technique used for event forecasting is graph modeling. Feng Chen et.al. \cite{Chen:2014:NSS:2623330.2623619} uses a heterogeneous graph modeling as keyword enrichment and pre-processing. To achieve accurate results in event forecasting they use only filtered entities for NPHGS approach. According to Feng Chen et.al. \cite{Chen:2014:NSS:2623330.2623619}, a heterogeneous graph is defined as a network consisting of nodes, edges and relations where nodes are the entities extracted using named entity recognition. There can be multiple types of nodes equivalent to number of entities extracted (topic, temporal, spatial, organization etc). Edges are the link between two entities and relation defines the feature vector between two entities. For example, one relation between a Twitter user $U$ (entity: people) and a term $T$ (entity: topic) can be number of tweets by user $U$ on topic $T$. Entities having high relevance and polarity score above certain threshold are filtered and used in next phase of NPHSG.\\
\indent Unlike Figure \ref{techniques_unrest}, due to large number of papers in online radicalization domain, we present machine learning technique over a timeline instead of individual presentation for each paper. Figure \ref{techniques_radicalization} reveals that text classification {KNN, Naive Bayes, Support Vector Machine, Rule Based Classifier, Decision Tree}, Clustering (Blog Spider), Exploratory Data Analysis (EDA), Topical Crawler/Link Analysis (Breadth First Search, Depth First Search, Best First Search) and Keyword Based Flagging (KBF) are the most widely used techniques for online radicalization detection on social media websites \cite{5137295} \cite{xu2006mining} \cite{agarwal2015using} \cite{6284099}. For all techniques, there are different set of discriminatory features which we will discuss in detail in Section \ref{section_features}. Social networking websites, micro-blogging websites and video sharing websites are amongst the largest repositories of user generated content on web. Therefore, Text classification (automatic and semi-supervised learning), clustering (unsupervised learning), EDA and KBF approaches are very well known techniques and commonly used for identifying extremist content on social media \cite{botha2013assessing} \cite{mahmood2012online}. While Topical crawler and Link analysis are the techniques used for crawling through navigation links and identifying similar users and locating hidden communities on social media websites \cite{procmicroposts2015@www2015} \cite{fu2010focused} \cite{agarwal2015topic}. The topical crawler is a recursive process that adds and removes nodes after each iteration. It starts from a seed node, traverses in a graph navigating through some links and returns all relevant nodes to a given topic. Breadth First Search, Depth First Search and Best First Search are different ways to pick neighbors and navigate through external links. These links and neighbors are different for different social networking websites. For example, if user $u$ posted a tweet $t$ then a neighbor can be a follower of user $u$, or users liking and re-tweeting or re-blogging (in case of Tumblr) that post. Similarly in YouTube, if user $u$ posted a video $v$ then a link can re-direct to a user channel who subscribed $u$ or posted a comment on video $v$.\\
\indent Language modeling, n-gram, Boosting are other techniques for classifying textual data as hate promoting based upon several discriminatory features \cite{agarwal2014focused}. TREC is a document ranking algorithm to search and rank documents available at Text REtrieval Conference. Bermingham et. al. \cite{bermingham2009combining} uses TREC algorithm to perform sentiment analysis of comments posted on extremist videos on YouTube. However, they find that due to the difference in nature of data, TREC is not useful to detect subjectivity in YouTube comments and makes the results unreliable. Since comments posted on YouTube are opinion based while the documents in TREC corpus are based upon facts.\\
\begin{table*}[t]
\centering
\caption{List of Various Dimensions and their Associated Properties Used for Annotating Literature Articles Conducted in the Domain of Civil Unrest Related Event Forecasting}\label{label_unrest}
\begin{tabular}{llp{0.65\textwidth}}
\hline
\textbf{Dimension}&\textbf{Categories}&\textbf{Description}\\
\hline
\multirow{5}{*}{\textbf{Features}}&F1 \quad Temporal&Presence of time related expressions\\
&F2 \quad Spatial&Presence of location based expressions\\
&F3 \quad Topic & Presence of targeted topic or domain related expressions\\
&F4 \quad Content & Mining text to extract event related information\\
&F5 \quad Demographic & Other demographic and statistical based metadata\\
\hline
\multirow{4}{*}{\textbf{Evaluation}}&E1 \quad K-Cross Validation&optimizing the output of forecasting by splitting data into k-samples\\
&E2 \quad Precision&Evaluates the exactness of forecasting results\\
&E3 \quad Recall & Evaluates the completeness of results\\
&E4 \quad NA & No evaluation technique is mentioned\\
\hline
\multirow{2}{*}{\textbf{Analysis}}&A1 \quad Content&Mining only textual content for feature extraction and event forecasting\\
&A2 \quad Community&Mining user profiles and networks for extracting event related information\\
\hline
\multirow{3}{*}{\textbf{Language}}&L1 \quad English&Conducting experiments only on English Language Tweets\\
&L2 \quad Non-English&Posts consisting of any Non-English language content\\
&L3 \quad Others&Testbed consisting of multiple languages content\\
\hline
\multirow{2}{*}{\textbf{Genre}}&G1 \quad Protest-Country Specific&Conducting Experiments for the protests happened in one or specific countries/region [Latin America]\\
&G2 \quad Global&Performing Event forecasting on any civil unrest related event happened worldwide\\
\hline
\end{tabular}
\end{table*}
\begin{table*}
\centering
\caption{Summary of Existing Literature in the Domain of Civil Unrest Related Event Forecasting}\label{summary_unrest}
\begin{tabular}{lccccccccccccccccc}
\hline
\multirow{2}{*}{\textbf{\hspace{1cm}Authors}}&\multirow{2}{*}{\textbf{Year}}&\multicolumn{2}{c}{\textbf{Genre}}&\multicolumn{2}{c}{\textbf{Analysis}}&\multicolumn{4}{c}{\textbf{Evaluation}}&\multicolumn{3}{c}{\textbf{Language}}&\multicolumn{5}{c}{\textbf{Features}}\\
&&\multicolumn{1}{c}{G1}&\multicolumn{1}{c}{G2}&\multicolumn{1}{c}{A1}&\multicolumn{1}{c}{A2}&\multicolumn{1}{c}{E1}&\multicolumn{1}{c}{E2}&\multicolumn{1}{c}{E3}&\multicolumn{1}{c}{E4}&\multicolumn{1}{c}{L1}&\multicolumn{1}{c}{L2}&\multicolumn{1}{c}{L3}&\multicolumn{1}{c}{F1}&\multicolumn{1}{c}{F2}&\multicolumn{1}{c}{F3}&\multicolumn{1}{c}{F4}&\multicolumn{1}{c}{F5}\\
\hline
Colbaugh et. al.&2010&\multicolumn{1}{c}{\Checkmark}&\multicolumn{1}{c}{\Checkmark}&\multicolumn{1}{c}{\Checkmark}&\multicolumn{1}{c}{\Checkmark}&\multicolumn{1}{c}{\Checkmark}&\multicolumn{1}{c}{}&\multicolumn{1}{c}{}&\multicolumn{1}{c}{}&\multicolumn{1}{c}{\Checkmark}&\multicolumn{1}{c}{}&\multicolumn{1}{c}{}&\multicolumn{1}{c}{\Checkmark}&\multicolumn{1}{c}{}&\multicolumn{1}{c}{}&\multicolumn{1}{c}{}&\multicolumn{1}{c}{}\\
\hline
Hua et. al.&2013&\multicolumn{1}{c}{\Checkmark}&\multicolumn{1}{c}{}&\multicolumn{1}{c}{\Checkmark}&\multicolumn{1}{c}{}&\multicolumn{1}{c}{}&\multicolumn{1}{c}{}&\multicolumn{1}{c}{}&\multicolumn{1}{c}{\Checkmark}&\multicolumn{1}{c}{}&\multicolumn{1}{c}{\Checkmark}&\multicolumn{1}{c}{}&\multicolumn{1}{c}{\Checkmark}&\multicolumn{1}{c}{\Checkmark}&\multicolumn{1}{c}{}&\multicolumn{1}{c}{\Checkmark}&\multicolumn{1}{c}{}\\
\hline
Ramakrishnan et. al.&2014&\multicolumn{1}{c}{\Checkmark}&\multicolumn{1}{c}{}&\multicolumn{1}{c}{\Checkmark}&\multicolumn{1}{c}{\Checkmark}&\multicolumn{1}{c}{}&\multicolumn{1}{c}{\Checkmark}&\multicolumn{1}{c}{\Checkmark}&\multicolumn{1}{c}{}&\multicolumn{1}{c}{\Checkmark}&\multicolumn{1}{c}{}&\multicolumn{1}{c}{\Checkmark}&\multicolumn{1}{c}{\Checkmark}&\multicolumn{1}{c}{\Checkmark}&\multicolumn{1}{c}{\Checkmark}&\multicolumn{1}{c}{\Checkmark}&\multicolumn{1}{c}{}\\
\hline
Jiejun et. al.&2014&\multicolumn{1}{c}{\Checkmark}&\multicolumn{1}{c}{}&\multicolumn{1}{c}{\Checkmark}&\multicolumn{1}{c}{}&\multicolumn{1}{c}{}&\multicolumn{1}{c}{\Checkmark}&\multicolumn{1}{c}{}&\multicolumn{1}{c}{}&\multicolumn{1}{c}{\Checkmark}&\multicolumn{1}{c}{}&\multicolumn{1}{c}{\Checkmark}&\multicolumn{1}{c}{\Checkmark}&\multicolumn{1}{c}{\Checkmark}&\multicolumn{1}{c}{}&\multicolumn{1}{c}{\Checkmark}&\multicolumn{1}{c}{}\\
\hline
Chen et. al.&2014&\multicolumn{1}{c}{}&\multicolumn{1}{c}{\Checkmark}&\multicolumn{1}{c}{\Checkmark}&\multicolumn{1}{c}{\Checkmark}&\multicolumn{1}{c}{}&\multicolumn{1}{c}{\Checkmark}&\multicolumn{1}{c}{}&\multicolumn{1}{c}{}&\multicolumn{1}{c}{\Checkmark}&\multicolumn{1}{c}{}&\multicolumn{1}{c}{}&\multicolumn{1}{c}{\Checkmark}&\multicolumn{1}{c}{\Checkmark}&\multicolumn{1}{c}{}&\multicolumn{1}{c}{\Checkmark}&\multicolumn{1}{c}{\Checkmark}\\
\hline
Compton et. al.&2014&\multicolumn{1}{c}{\Checkmark}&\multicolumn{1}{c}{}&\multicolumn{1}{c}{\Checkmark}&\multicolumn{1}{c}{}&\multicolumn{1}{c}{}&\multicolumn{1}{c}{\Checkmark}&\multicolumn{1}{c}{}&\multicolumn{1}{c}{}&\multicolumn{1}{c}{\Checkmark}&\multicolumn{1}{c}{}&\multicolumn{1}{c}{\Checkmark}&\multicolumn{1}{c}{\Checkmark}&\multicolumn{1}{c}{\Checkmark}&\multicolumn{1}{c}{}&\multicolumn{1}{c}{\Checkmark}&\multicolumn{1}{c}{\Checkmark}\\
\hline
Filchenkov et. al.&2014&\multicolumn{1}{c}{}&\multicolumn{1}{c}{\Checkmark}&\multicolumn{1}{c}{\Checkmark}&\multicolumn{1}{c}{}&\multicolumn{1}{c}{}&\multicolumn{1}{c}{}&\multicolumn{1}{c}{}&\multicolumn{1}{c}{\Checkmark}&\multicolumn{1}{c}{\Checkmark}&\multicolumn{1}{c}{}&\multicolumn{1}{c}{}&\multicolumn{1}{c}{\Checkmark}&\multicolumn{1}{c}{}&\multicolumn{1}{c}{}&\multicolumn{1}{c}{}&\multicolumn{1}{c}{}\\
\hline
Muthiah et. al.&2015&\multicolumn{1}{c}{\Checkmark}&\multicolumn{1}{c}{}&\multicolumn{1}{c}{\Checkmark}&\multicolumn{1}{c}{\Checkmark}&\multicolumn{1}{c}{}&\multicolumn{1}{c}{\Checkmark}&\multicolumn{1}{c}{\Checkmark}&\multicolumn{1}{c}{}&\multicolumn{1}{c}{\Checkmark}&\multicolumn{1}{c}{}&\multicolumn{1}{c}{\Checkmark}&\multicolumn{1}{c}{\Checkmark}&\multicolumn{1}{c}{\Checkmark}&\multicolumn{1}{c}{}&\multicolumn{1}{c}{\Checkmark}&\multicolumn{1}{c}{}\\
\hline
\end{tabular}
\end{table*}
\indent OSLOM (Order Statistics Local Optimization Method) is a clustering algorithm designed for graphs and networks which has been used by some researchers to locate groups and communities of extremist users sharing a common agenda. OSLOM is an open source visualization tool capable to detect hidden communities in a network accounting for edge directions, edge weights, overlapping communities, hierarchies and community dynamics \cite{o2013uncovering} \cite{o2013analysis}. OSLOM locally optimize the clusters and consists of following three phases: 
\begin{enumerate}
\item Iterative process to identify significant clusters of nodes
\item Analyzing resultant clusters and identification of internal structure and overlapping clusters (if possible)
\item Identification of hierarchical structure of clusters.
\end{enumerate}
\subsection{Characterization and Classification of Articles- Early Warning and Prediction of Civil Unrest Related Events}\label{section_features}
Table \ref{label_unrest} illustrates various facets and their associated properties used by us for performing an in-depth characterization of existing literature. Table \ref{summary_unrest} shows the summary of this characterization performed on all $8$ papers. Features mentioned in the Table \ref{label_unrest} are mainly of three types: entities, content analysis and demographic information. Entities are extracted using named entity recognition (refer to Section \ref{Section-datamining-techniques}) and are pre-defined for domain specific problem. These entities are temporal expressions (today, tomorrow, $10$pm), spatial location expressions (USA, in front of white house, geocodes) and the topic being discussed in posts (migration, protest). Content analysis includes the mining and analyzing contextual metadata. For example, presence of hashtags, @user mentioned in a tweet. Demographic information are the statistical based measures. For example- number of tweets posted by a user, age and location of user. Table \ref{summary_unrest} reveals that $75$\% of papers have used spatiotemporal features as discriminatory features for predicting events. While rest of the $25$\% uses only timeline as a discriminatory features and uses event tracking for prediction. Table \ref{summary_unrest} also reveals that all the papers using spatiotemporal features also use content analysis to extract more relevant information from raw text. Content analysis feature includes searching for domain specific key terms.\\
\indent Precision, Recall and K-cross validation are standard information retrieval techniques to evaluate the performance of proposed techniques. Table \ref{label_unrest} reveals the common evaluation measures used by researchers. Precision computes the 'exactness' of proposed method while recall computes the 'completeness' of solution approach. K-cross validation splits the testbed into K parts and evaluates the model iteratively for each part. K-cross validation is used to optimize the results. Table \ref{summary_unrest} reveals that $65$\% of the studies use Precision measure to evaluate the performance of their proposed approach among which $25$\% papers use both recall and precision measures. While in $2$ papers no evaluation method is defined and in only $1$ paper, authors use k-cross validation method to measure the effectiveness of their approach.\\
\indent Right prediction of an event depends upon the content present in the documents, tweets and other contextual metadata. However, we observe that some researchers also use profile information of posters and groups of these posters talking about an event for event forecasting. Therefore, we categorize these papers based upon two classes of analysis: content and community. Content, where author mine only contextual metadata for feature extraction and event prediction. Community, where authors mine both user profiles and network (links) for extracting event related information. Table \ref{summary_unrest} reveals all papers have used content analysis for predicting events related to protest and civil disobedience. While in $50$\% of the articles, authors use profile information of users posting event related content and a bunch of users talking about the event. This information includes information about geographical location (geocodes), frequent '@' mentioned (to find other connected user profiles), frequent hashtags used (to expand the list of domain and event specific keywords).\\
\indent Popular social media website allow users to post content in multiple language being spoken across the world. Translating multi-lingual text into English language text and extracting information from it is useful when predicting a country or region specific events. We classify these papers into $3$ categories based upon the language being addressed for event prediction. Table \ref{summary_unrest} reveals that in $90$\% of the papers authors have worked on English language text among which in $60$\% the papers they have addressed multiple languages (both English and Non-English) text. Dutch, French, Spanish and Portuguese are the languages that were addressed in maximum papers. We also find one paper where authors have worked on only Spanish language tweets.\\
\indent We classify articles into different genre based upon the targeted region defined in articles. Based upon our analysis we define two genre for region based classification: Country Specific (example- Latin America, Africa) and Global. Table \ref{summary_unrest} reveals that $75$\% of papers focus on events specific to a region or a country among which only one article addressing the events happening across all local and global regions.
\begin{table*}
\centering
\caption{List of Various Dimensions and their Associated Properties Used for Annotating Literature Articles Conducted in the Domain of Online Radicalization Detection}\label{label_radicalization}
\begin{tabular}{llp{0.65\textwidth}}
\hline
\textbf{Dimension}&\textbf{Categories}&\textbf{Description}\\
\hline
\multirow{3}{*}{\textbf{Features}}&F1 \quad Text&Textual based content of the posts (Title of Video, Tweet, Facebook Comments etc.)\\
&F2 \quad Link&Links between two user profile (Subscription in YouTube, Follower in Twitter and Tumblr etc.)\\
&F3 \quad Demographic&Other demographic and statistical based metadata\\
\hline
\multirow{8}{*}{\textbf{Evaluation}}&E1 \quad Precision&Evaluates the exactness of forecasting results\\
&E2 \quad Recall & Evaluates the completeness of results\\
&E3 \quad F-Score & Weighted harmonic mean of Precision (E1) and Recall (E2)\\
&E4 \quad Accuracy & Evaluates the correctness of the technique\\
&E5 \quad K-cross Validation & optimizing the output of forecasting by splitting data into k-samples\\
&E6 \quad SNA & Performs Social Network Analysis to show the findings in community detection\\
&E7 \quad User Based & Evaluation performed by external users\\
&E8 \quad NA & No evaluation method is defined.\\
\hline
\multirow{3}{*}{\textbf{Analysis}}&A1 \quad Content&Mining only textual content for feature extraction and event forecasting\\
&A2 \quad User Profile&Mining user profiles metadata for extracting event related information\\
&A3 \quad Community&Mining linked profiles and their communities\\
\hline
\multirow{3}{*}{\textbf{Language}}&L1 \quad English&Conducting experiments only on English Language Tweets\\
&L2 \quad Arabic&Testbed consisting of content written in Arabic language (scripted or language text)\\
&L3 \quad Other Non-English&Posts consisting of any other Non-English language (excluding L2) content (German, French etc.)\\
\hline
\multirow{3}{*}{\textbf{Region}}&R1 \quad US Domestic&Content posted targeting US issues or radicalization originated from US domestic regions\\
&R2 \quad International&Researchers focusing on global or international radicalization or extremism\\
&R3 \quad Others&Radicalization originating or targeting other groups and regions worldwide (example- Middle Eastern, Latin America)\\
\hline
\multirow{5}{*}{\textbf{Genre}}&G1 \quad Anti-black&Content posted by white supremacy communities targeting black people\\
&G2 \quad Jihad&Groups posting content for promoting Jihad among their viewers\\
&G3 \quad Terrorism&Content posted by terrorism group or social networking activity performed by terrorists\\
&G4 \quad Hate and Extremism&Content posted in order to promote hate and extremism among various targeted audience\\
&G5 \quad Religion&Content posted against a religion (example- Anti-Islamic Tweets)\\
\hline
\end{tabular}
\end{table*}
\subsection{Characterization and Classification of Articles- Identification of Extremist Content, Users and Hidden Communities}
\begin{figure*}[t]
\centering
\begin{minipage}{.50\textwidth}
 \centering
 \includegraphics[scale=0.05]{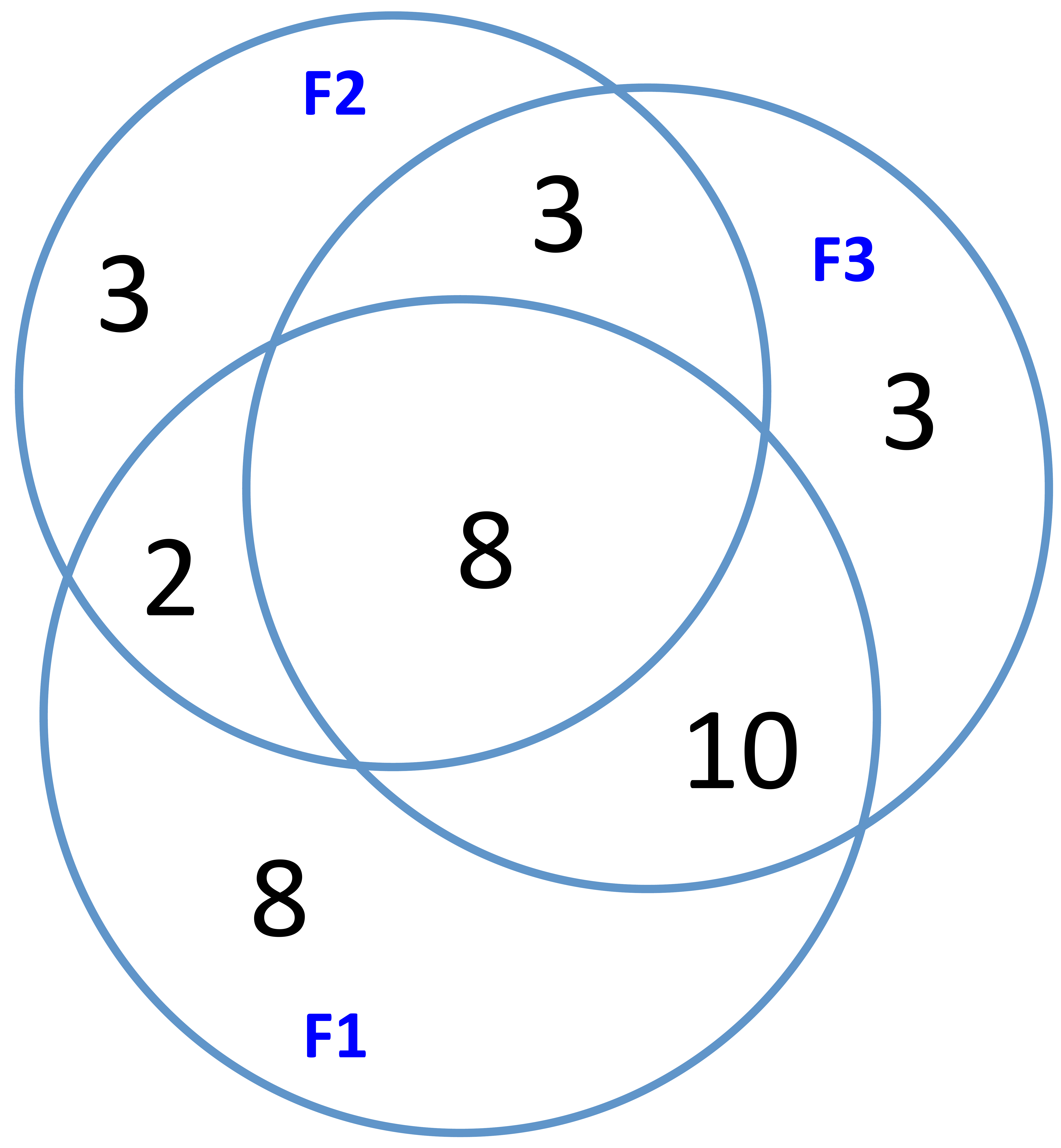}
 \captionof{figure}{Number of Articles Classified Based upon the Discriminatory Features Used for Online Radicalization Detection}
 \label{fig:features}
\end{minipage}%
\begin{minipage}{.50\textwidth}
 \centering
 \includegraphics[scale=0.05]{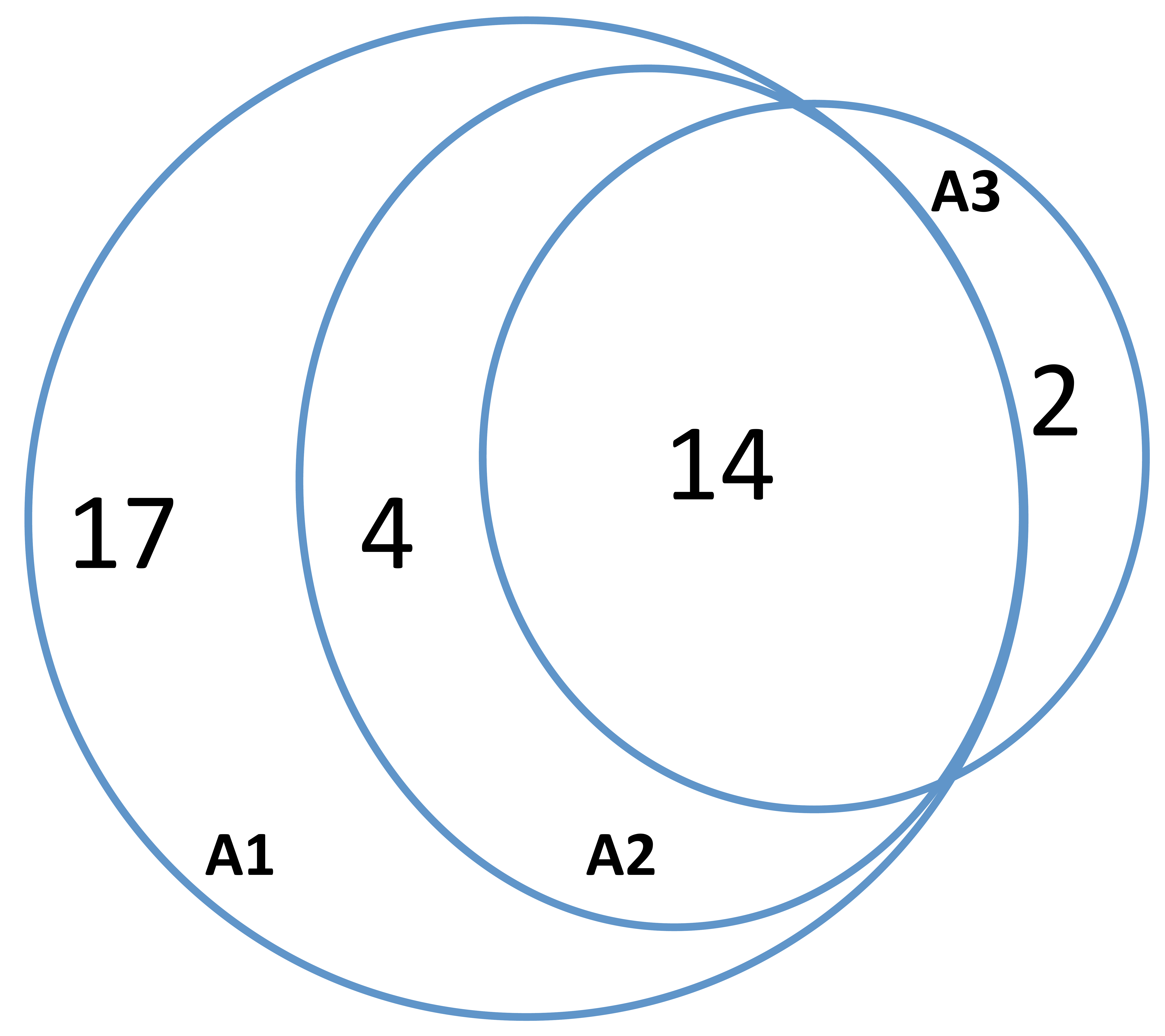}
 \captionof{figure}{Number of Articles Classified Based upon the Metadata Being Used for Extremism Detection Methods}
 \label{fig:analysis}
\end{minipage}%
\end{figure*}

\indent Table \ref{label_radicalization} shows various dimensions and their associated categories used by us to classify existing scholarly articles in the domain of online radicalization detection.\\
\indent In online radicalization detection literature review, we collect articles focusing towards identification of extremism content and locating hate promoting users and communities. Therefore we classify these articles based upon three discriminatory features being used by researchers. Text defines the contextual metadata of a post. For example, content of a tweet, title of a video, tags present in the description of a video etc. Link defines the relation between two user profile which can be different for different social media platforms. For example, "post re-blogged by" in Tumblr, a "follower" in Twitter, "video liked by" in YouTube etc. Demographic information are the statistical based metadata of post and user profiles. For example, number of comments on a video, number of favorites on a tweets, notes on a Tumblr post etc. Figure \ref{fig:analysis} shows the statistics of these features used by various articles collected in the domain of online radicalization. Figure \ref{fig:analysis} reveals that the number of papers using text, link and demographic information are $16$, $24$ and $28$ respectively. We notice that $25$\% of these papers use all three features while $30$\% of these papers use contextual and demographic information for extremist content detection. There are very few papers using only link information and statistical data for online radicalization detection.\\
\indent Similar to Table \ref{label_unrest}, we classify articles into $3$ sub-categories i.e. 'content', 'user profile' and 'community' based upon the type of content being analyzed for providing solutions to counter and combat online extremism. Figure \ref{fig:analysis} shows a Venn diagram for number of articles classified in each category. Figure \ref{fig:analysis} reveals that out of $37$ articles, in $35$ papers, authors conduct experiments on contextual metadata while in $18$ papers, authors extract user profile information. Similarly, in $16$ articles, they mine linked user profiles and their communities.\\
\begin{figure*}[t]
\centering
\begin{minipage}{.50\textwidth}
 \centering
 \includegraphics[scale=0.05]{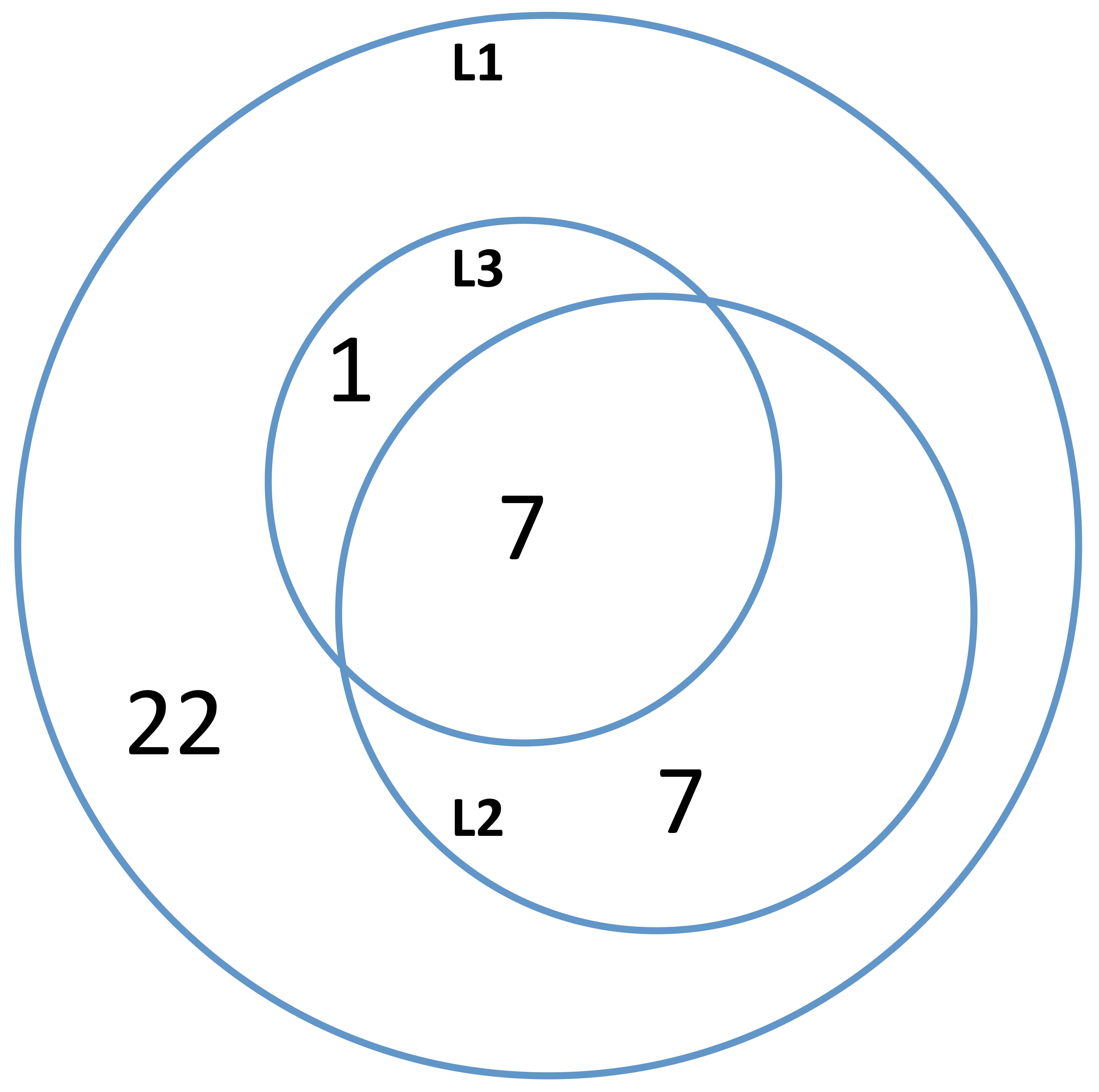}
 \captionof{figure}{Number of Articles Classified Based upon the Language Being Addressed in Proposed Methodology}
 \label{fig:language}
\end{minipage}%
\begin{minipage}{.50\textwidth}
 \centering
 \includegraphics[scale=0.05]{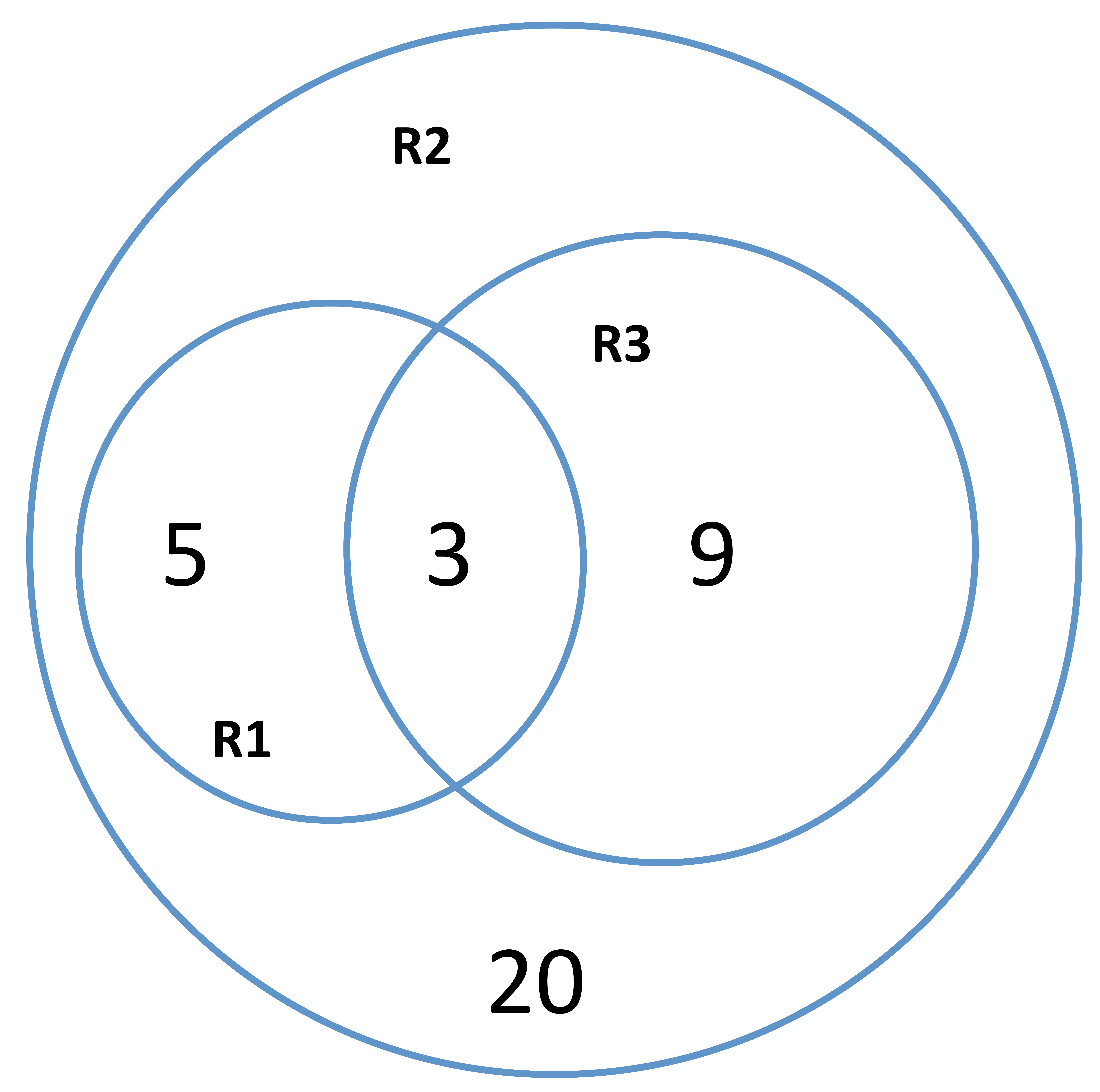}
 \captionof{figure}{Number of Papers Classified Based upon the Targeting and Originating Region of Online Radicalization}
 \label{fig:region}
\end{minipage}
\end{figure*}
\indent As shown in Figure \ref{fig:analysis}, content analysis is an important feature for identifying extremist content. Presence of multi-lingual text in contextual metadata makes it technically challenging to mine and analyze the data. Based upon our analysis, we classify articles into three categories based upon the language being addressed in experimental setup. Figure \ref{fig:language} shows that among $37$ articles, $14$ articles are capable to mine and identify extremist content posted in Arabic language. Among these $14$ articles, methods proposed in $7$ articles are capable to analyze other non-English language texts as well. Figure \ref{fig:language} also reveals that all proposed methodologies in previous researches are capable to detect extremist content posted in English language.\\
\begin{figure*}[t]
\centering
\includegraphics[scale=0.50]{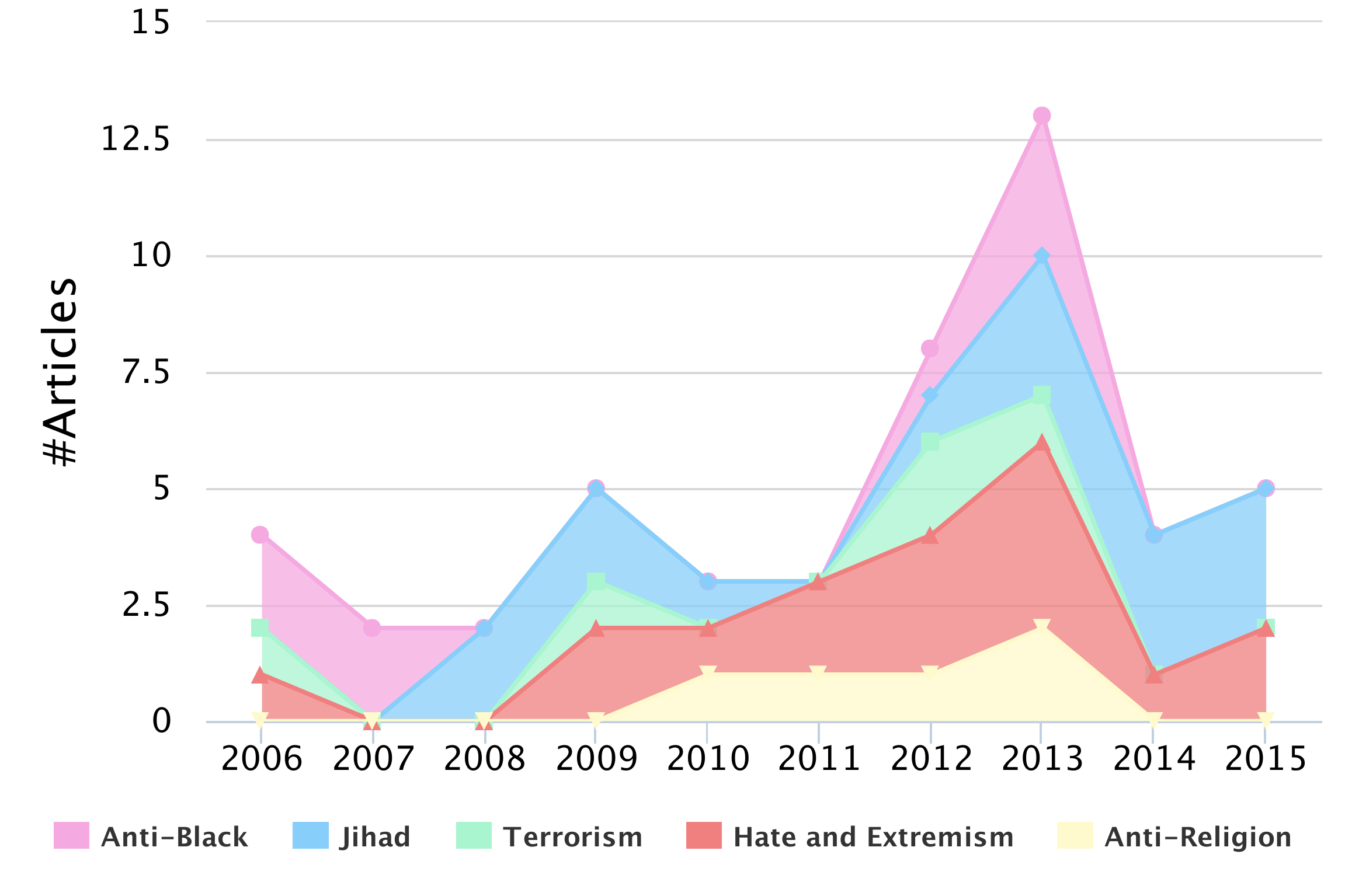}
\caption{Distribution of Number of Publication over Past Decade Targeting Several Domains of Online Radicalization}
\label{genre}
\end{figure*}
\begin{figure*}[t]
\centering
\includegraphics[scale=0.50]{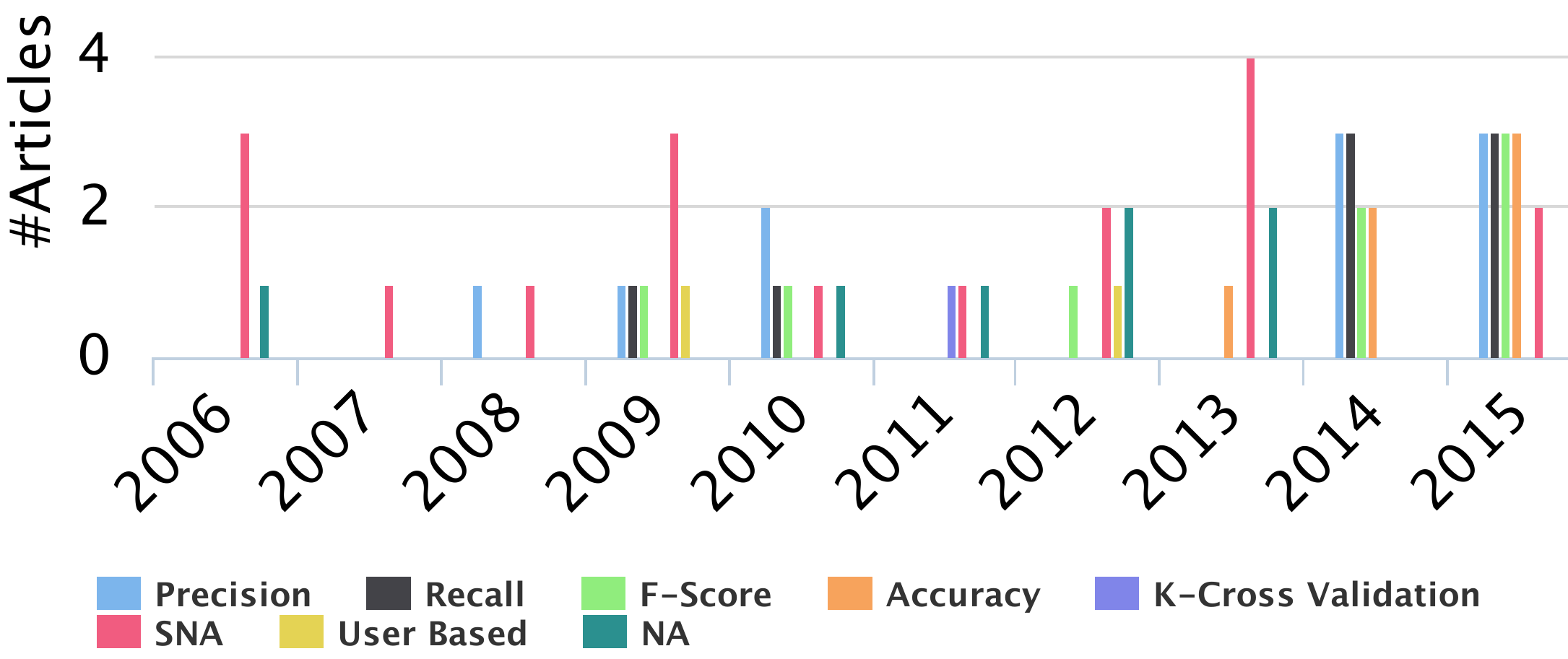}
\caption{Distribution of Number of Articles over past 10 Years using Different Evaluation Techniques}
\label{evaluation}
\end{figure*}
\indent As discussed in Section \ref{section_features}, many of the articles in online radicalization have been published in counter-terrorism and deradicalizing extremism. The focus of these researches is to counter and combat online radicalization targeting a specific country and originated from a specific region. Based upon the target region and country we classify these $37$ articles into three categories: US Domestic, International and Others. Figure \ref{fig:region} shows the Venn diagram of number of articles classified in each category. Figure \ref{fig:region} reveals that there are $20$ articles focusing on countering online radicalization happening worldwide. $5$ articles focus on extremist content targeting issues of US domestic and originated from US domestic regions. In existing literature we find $9$ such articles that focus on other extremist groups and regions worldwide. For example, North Africa, Latin America, Middle Eastern countries. We also find $3$ articles targeting both US domestic, Middle Eastern and other regions for detecting online radicalization.\\
\indent There are various genre of radicalization present on Internet. Based upon previous researches we divide genre into $5$ categories: Anti-black or white supremacy communities are a group of people targeting non-whites. Anti-black is a form of racism spreading and promoting the ideology that white people are superior in certain characteristics, traits, and attributes to other non-white people. Jihad or Islamic extremism communities are the groups promoting Islamic extremism on social media. Act of Islamic extremism includes promoting "Sharia law" and abusing human rights and terrorism (However, here we use terrorism as an independent category for classification). Anti-religion radicalization is posting hateful speech against a religion and promoting the beliefs and ideology of oneself. Hate and Extremism includes the others categories that are undefined and covers political radicalization. Figure \ref{genre} shows the number of articles classified in each genre. Due to the large number of articles in online radicalization domain, we use timeline on x-axis and number of articles on y-axis. Figure \ref{genre} reveals that over past decade there has been a constant research in each category of online radicalization. However, we observe that through out this timeline maximum number of publications are in the category of Jihad and terrorism. In early 20s there has been some work in the domain of white supremacy community detection on social media but after $2008$, we find a sudden increase in the number of publication focusing on anti-black communities ($13$ publications). Figure \ref{genre} also reveals that the number of publications focusing on anti-religion based content are very few in comparison to other categories of online radicalization.\\
\indent As discussed in previous Section (\ref{section_features}), Precision, Recall, K-cross validation are standard information retrieval measures to evaluate the effectiveness of proposed solution approach. Based upon the literature articles of online radicalization detection we divide Evaluation dimension into $8$ subcategories (illustrated in Table \ref{label_radicalization}). F-score evaluates the harmonic mean of Precision and Recall while Accuracy computes the "correctness" of proposed method. Social Network Analysis (SNA) is an intuitive based technique to uncover patterns and relationships between entities in a network. User based analysis are the evaluations performed by external users. Figure \ref{evaluation} demonstrates the distribution of number of articles using various measures to examine the performance of proposed methods over past decade. As shown in Figure \ref{fig:analysis}, $50$\% of the articles uncover the hidden communities of hate and extremist users. We find a similar pattern in Figure \ref{evaluation} which reveals that Social network analysis is the most commonly used technique to evaluate the performance of proposed method. Figure \ref{evaluation} also reveals that there is only one article using K-cross validation for evaluation while in $7$ articles no evaluation method is defined. We also observe that despite having $50$\% of the articles that analyze only textual data to identify extremist content, only $30$\% of the articles use precision while only $20$\% of the articles use recall as a measure for evaluation.
\section{Closing Remarks}
\indent Applying social media intelligence for predicting and identifying online radicalization and civil unrest oriented threats is an area that has attracted several researchers' attention over past $10$ years. We observe a surge in research interest over the last $3$ years on the topic of solutions for identifying and forecasting civil unrest and mobilization by mining textual content in open-source social media. The number of research papers on social media analytics for online radicalization detection are much more (about $7$ times) than on civil disobedience detection. Intelligence and Security Informatics (ISI) conference and Security Informatics (SI) journal are the two main venues for publishing papers on the topic of online radicalization detection and mining. Our analysis reveals that micro-blogging websites like Twitter and Tumblr are the two most common sources of social media data for civil unrest detection and forecasting applications. We believe that Twitter has been very instrumental in facilitating political mobilization in comparison to other social media platforms because of its inherent characteristics of sharing short text through direct messages and follower relationship. It is interesting to observe that despite the immense popularity and penetration of YouTube as an online video-sharing website, it has not been used in any of the existing research for protest planning or prediction. On the contrary, our research reveals that YouTube is the most widely used platform for online radicalization, hate and extremism promotion as indicated by the published research papers. In comparison to Twitter, Tumblr which is also a popular micro-blogging website has not been a major focus of research attention for online radicalization detection applications.\\
\indent Our analysis reveals a variety of information retrieval and machine-learning based methods and techniques used by researchers to investigate solutions for civil unrest and online radicalization detection. Clustering, Logistic Regression and Dynamic Query Expansion are the commonly used techniques to predict upcoming events related to civil unrest or protest. We observe that Named Entity Recognition (NER) is a common component in the text processing pipeline for various proposed approaches and techniques. Graph modeling is also a technique adopted by several researchers for the problem of event forecasting. Our survey reveals that KNN (K Nearest Neighbor), Naive Bayes, Support Vector Machine, Rule Based Classier, Decision Tree, Clustering (Blog Spider), Exploratory Data Analysis (EDA), Topical Crawler/Link Analysis (Breadth First Search, Depth First Search, Best First Search) and Keyword Based Flagging (KBF) are the most widely used techniques for online radicalization detection on social media websites.\\
\indent In this survey, we categorize existing studies on various dimensions such as the use of discriminatory features, type of metadata being analyzed and evaluation techniques used by authors to examine the effectiveness of their approach. Our analysis reveals mining contextual metadata of a post and spatiotemporal information present in the content are most commonly used feature for predicting civil unrest related events. In existing studies, we find that maximum researchers evaluates the accuracy of their prediction approach by computing precision of their results. We also observe that $90$\% of the studies are able to mine English language text. Since the problem of civil unrest related event prediction can be targeted to a specific country or region, we find that the methods proposed in various studies are able to mine information from multi-lingual texts (Dutch, Spanish, French etc). Our analysis also reveals that $60$\% of the studies targets events specific to a country ot region. Maximum number of studies are conducted on the events happened in Latin America and USA.\\
\indent Characterization and meta analysis performed on the existing studies of online radicalization detection reveals that to identify the presence of extremist content contextual based metadata is most commonly used feature. However, demographic information and activity feeds of a user profile and links between two users are discriminatory features for locating hidden communities of extremist users. We observe that many of the existing techniques are capable to mine multi-lingual text such as Arabic and capture relevant information. Our analysis reveals that there are papers that targets only a country and region specific radicalization (Latin America, Middle Eastern, North Africa). We also observe that to examine the effectiveness of results in community detection and extremism content identification, Social Network Analysis and Precision measures are most commonly used evaluation methods.


\begin{backmatter}




\bibliographystyle{bmc-mathphys} 
\bibliography{bmc_article} 
\nocite{*}



\end{backmatter}
\end{document}